\documentstyle[12pt] {article}
\title{The Very Early Universe}
\author{V.N.Lukash\\
Astro Space Center of Lebedev Physical Institute\\
Profsoyuznaya, 84/32, 117810 Moscow, Russia}
\begin{document}
\maketitle

In these lectures \footnote{Published in: ''Cosmology: The Physics of 
the Universe'', eds B.A.Robson et al., World Scientific, 213 (1996), and 
``Cosmology and Gravitation, II'', ed. M.Novello, Editions Frontieres, 288 
(1996)} we dwell upon the cosmological corner-stones of the Very
Early Universe (VEU) theory: Parametric Amplification Effect (PAE) 
responsible for the generation of Primordial Cosmological Perturbations 
(PCPs), Chaotic and Stochastic Inflation, Principal Tests of VEU, and others.

\section{Introduction}
A great success of VEU theory as the theory of the beginning of the Big Bang, 
is related to its semiclassical nature allowing to operate productively in
terms of classical space-time filled with quantum physical fields (including 
the gravitational perturbations). It (VEU theory) connects like a bridge
the theory of our Universe based on the Friedmann model (FU), with
theories of Everything (TOE) essentially employing quantized gravity
(still very ambiguous). This connection is already realized itself in
the important understanding that the quasi-homogeneous and isotropic
state of the Universe on the horizon scale (on one side) and the
primordial cosmological perturbations which gave a birth to the
Universe structure on smaller scales (on the other side) are just two
features in the low-energy limit of some theory of VEU based on a
model of the inflationary Universe (IU). Up to now we have no
alternative to the Inflation resulting in the cold remnants which we
observe today as the micro and macro worlds, which is commonly considered as
a necessary element and the probe test of high energy physics and any TOE.  

We should begin from the priorities when discussing VEU theory:
whether to start from cosmological or particle physics standards. The
particle physics is not yet fixed well at high energies: to follow this
direction today means to start from  N particle and modified gravity
theories where N is a big number, and then to build up N inflationary
models based on them. For this reason, rather preferable now seems the
investigation of cosmologically standard VEU  theories based of three
standard points: 
\begin{itemize}
\item[(i)] setting some cosmological postulates within General Relativity 
(GR);
\item[(ii)] deriving the theory from these postulates, and 
\item [(iii)] confronting the theory predictions with observations.
\end{itemize}

On this way, which I will follow in these lectures, we are independent
of the future particle physics and, thus, may try to find some
basic properties and principal features of the VEU theory (tested by
observations!) which stay independent of the particle physics
uncertainties as well. We know two good examples of such kind
theories. The first is the Friedmann-Robertson-Wolkel model which has
ensured the great success of the observational and theoretical
cosmology, just based on the cosmological postulate (the  homogeneity
and isotropy of the 3-space). The second is PAE, the theory of
generation of PCPs which gave birth to the Large Scale Structure (LSS)
of the Universe, just based on the linear perturbation theory in FU.

Regarding the VEU our goal would be to find the basics for IU in GR which 
could create the Friedmannian region we live in now (which, in its turn, 
would provide for the initial conditions for both our examples, the FU and
PAE theories). The answer which we know today is the model of Chaotic 
Inflation based on the assumptions of existence of inflaton (the scalar 
field weakly coupled to all other physical field) and 
the start-inflation-condition postulating a quasi-homogeneous spatial 
distribution of the initial inflaton in a Compton-wavelength region 
(i.e., in the finite-scale region! A great step in comparison with the FU 
cosmological postulate !).

Below, our lectures will be grouped in four parts. The first is
devoted to the PAE  of scalar perturbations in FU (Lukash 1980). Next, 
we consider the cosmological applications to the {\it scattering} problem 
and chaotic inflation (Lukash $\&$ Novikov 1992). In the third part 
we present some recent developments of the theory of chaotic and
stochastic inflation proposed by Linde (1983, 1986). The last Chapter
deals with the problem of testing and confronting these theories with 
observations.  

Certainly, we do not pretend to cover all the corresponding references
and give a review of all recent ideas and speculations in VEU theories.
Our main goal here is to present some basic properties of VEU which
seem today more or less settled and independent of future  theoretical
constructions. We try to consider the simplest mathematic models paying
particular attention to the physical meaning of the effects considered.
Some necessary mathematical calculations are given in three Appendices.
Hereafter, our units imply $c=8\pi G=\hbar=1$ and $H_{o}=100 h\,km\, 
s^{-1} Mpc^{-1}$.

\section{Parametric Amplification Effect}
The formation of the structure of the Universe is one of the
fundamental problems of the modern Cosmology. The two following
properties of the present Universe are very important in understanding
the physics of its early expansion.
\begin{itemize}
\item [(i)] High degree of homogeneity and isotropy at large scales ( $\delta
\rho/ \rho {}^{<}_{\sim} 10^{-4}$ on the cosmological horizon) along
with a well-developed structure on scales less than 0.01 of the horizon.
\item [(ii)] High specific entropy $(N_{\gamma} /N_{b} \simeq 10^{9})$ along
with the mean baryon density being less than $\sim 10 \%$ of the critic
value $(\Omega_{b} {}^{<}_{\sim} 0.1)$.
\end{itemize}

Property (i) proves that the large scale structure of the Universe
(galaxies, clusters and supercluster) stemmed from initially small
amplitude perturbations of homogeneous and isotropic cosmic medium
since it is the small perturbation that may grow up to the order of
unity (and then form gravitationally bounded object) only when the
horizon becomes many times larger than its linear scale. We do not know
yet whether PCPs formed together with the cosmological Friedmann model
at Planckian curvatures of whether they originated in the process of
the homogeneous and isotropic expansion which is described by the
Classical General Relativity (CGR). In the first case we have no
theory. However, an important point is that the quasi-Friedmannian
Cauchy-hypersurface is already a classical object after Planckian time.
So, if the PCPs are made evolutionary and their scales less than the
Friedmann-hypersurface scale then they are likely to form at a
semiclassical stage when the large-scale gravity was governed by the CGR 
equations. We know one example of this kind - it is inflation: galaxy-scale
PCPs form at the very late stages of the inflationary expansion when
the Friedmannian Cauchy-hypersurface (which forms the background of our
local Universe today) has been already prepared by the inflation.
Below, we will develop the theory of small potential (scalar)
PCPs assuming the existence of Friedmann background model.

Also, it is quite evident that by no process could the Universe be born
strictly homogeneous and isotropic: there always exist quantum
fluctuations of metric and physical fields, the seed fluctuations could
be of statistical, random character or they might be thermal, etc. An
important point here is as follows: the inevitable minimum level of
seed initial perturbations is always maintained by the quantum
point-zero fluctuations of a quasi-homogeneous gravitating medium
which bases the Friedmann spatial slice.

One of the basic implications of property (ii) is that in the past the
Universe was hot and its expansion was governed by the gravitational
field of intensively interacting relativistic particles. At this stage
the matter represented to high degree of accuracy a hydrodynamic perfect 
fluid with equation of state $p=\epsilon/3$ (The Hot Friedmann Universe, HFU).

Small perturbations of such radiationally dominated gravitating fluid are 
sound waves propagating through the matter with a constant amplitude 
(adiabatic decrease of the wave amplitude due to the Universe expansion 
is exactly compensated in this situation by the increase of the amplitude
gained due to the pressure gradient in the comoving to the wave front
reference system). Mathematically, such perturbations in the expanding
Universe are governed by the same equation as that for the acoustic
perturbations of a non-gravitating homogeneous static termal bath in the flat
Minkowski space-time (conformal invariance). A deep physics is behind it:
no new phonons, cosmological potential perturbations, can be produced
during the expansion of the hot Universe.

For the real generation of PCPs to occur in the early Universe, one
has to reject the hot equation of state $(p = \epsilon/3)$ at some expansion
epoch, which can be done in principle only in VEU before the
primordial heating of cosmic matter to high relativistic
temperatures. In this case, as we shall see below, the number of
phonons is not conserved and new phonons (the density perturbations)
can be created in the course of the expansion. The mechanism is purely
classical and works as parametric amplification:  energy of large-scale
non-stationary gravitational potential is pumped to the energy of
small-scale perturbations (like new photons are created in an
electromagnetic resonator when its size changes in non-adiabatic
way). This effect, which we will generally call "parametric
amplification effect", has nothing  to do with Jeans instability:
actually it is the CGR-effect since the typical scales to be amplified are
just the cosmological horizon (and the effect involves the light velocity and
gravitational fundamental constants). Before we present the mathematical 
formalism the physical meaning of the parametric amplification is
discussed in the next Section. 

\subsection{Physical Meaning of the Parametric Amplification}
The theory of small perturbations in the FU was constructed by Lifshitz
in 1946. According to this theory there are three types of
perturbations of the homogeneous isotropic model: density (potential)
perturbations, vortex perturbations and gravitational waves. We are
interested now in the first type of perturbations, potential
perturbations, because they are related to galaxy formation.

Let us consider the spatially Euclidean background model:
$$
ds^2=dt^2-a^2d\vec{x}^2=a^2\left(d\eta^2-d\vec{x}^2\right), \eqno(1)
$$
where $a$ is the scale factor which is a function of time, $t$ and
$\eta$ are universal and conformal times respectively,
$$
\eta = \int \frac {dt}{a}  \eqno(2)
$$
It is usual to present the density perturbations $\delta \rho$ in a
Fourier expansion
$$
\delta = \frac{1}{(2\pi)^{3/2}}\int d^{3} \vec{k} \delta_{\vec{k}} e^{i
\vec{k} \vec{x}}, \eqno(3)
$$
where $\delta = \delta \rho /(\rho + p)$ (more rigorous definition
which makes $\delta = \delta (t, \vec{x})$ the  gauge invariant function is
given later on). Here, the wavenumber $k = |\vec{k}|$ and the
physical wavelength $\lambda = 2 \pi a/k. $

Also we shall use the perturbation scale
$$
l_k=\frac{\lambda}2 = \frac {\pi a}{k}, \eqno(4)
$$
and the horizon scale\footnote {For brevity, we refer to
the Hubble scale $H^{-1}$ as horizon, although it is not technically correct.}
$$
l_{H} = H^{-1} = \frac {a}{\dot{a}} \eqno(5)
$$
The latter is a typical scale of the causally connected region during
the evolution time scale of the $a$-function. (Dot is the derivative
over the universal time $(\dot{}) = \frac {d}{dt}).$

We consider here only homogeneous and isotropic  states of the
perturbation fields (random spatial phase fields). Their important
characteristic is a power spectrum of the density perturbations 
$\Delta^{2}_{k}$:
$$ 
\frac{k^{3}}{2 \pi^{2}} \langle \delta_{\vec{k}} \delta^{\ast}_{\vec{k}^{'}}
\rangle = \Delta^{2}_{k} \; \delta(\vec{k} - \vec{k}^{'}), \eqno(6)
$$
where brackets $\langle ... \rangle$ mean average over the field state,
$\delta(\vec{k} - \vec{k}^{'})$ is the 3-dimensional $\delta$-function and
$({}^{\ast})$ is the complex conjugate. This spectrum determines the
second correlation of the density perturbations
$$
\xi(r) = \langle \delta(t, \vec{x}) \delta (t, \vec{x} + \vec{r}) \rangle =
\int \limits^{\infty} _{o} \frac {dk}{k} \Delta^{2}_{k} \frac {\sin (kr)}{kr},
\eqno(7)
$$
where $r = |\vec r|$. All the odd correlations are identically zero.
The Gaussian fields which are the particular cases of the random
fields, can be totally described only by this correlation function (all
higher order correlations are negligible in linear approximation).
Eq. (7) also clarifies the physical sense of the power spectrum. The
amplitudes $\Delta_{k}$ are just the corresponding density perturbations in
the scale interval $\Delta k \sim k$ around $k$, they are additive (the
total power is a sum of the partial ones over all scales):
$$
\langle \delta^{2} \rangle = \sum_ {\Delta k \sim k} \Delta^{2}_{k}
$$

Let us now consider the hot Universe (HFU) when equation of state
was \footnote {In our units the energy density and matter density are
the same functions, $\epsilon = \rho c^{2} = \rho$.} $p =
\epsilon/3$. The solution for $\Delta_{k}$ is
$$
\Delta^{2}_{k}=(c_{1}(k)f_{1}(\kappa))^2+(c_2(k) f_2(\kappa))^2,\eqno(8)
$$
where
$$
\kappa = \omega \eta \sim \frac{l_{H}}{l_{k}}\sim \frac{t}{l_k}
\sim a \sim t^{1/2}, \,\, \omega = \frac {k}{\sqrt {3}},
$$
$$
f_{1}(\kappa) = - \cos \kappa + 2 \left (\frac {\sin \kappa}{\kappa} + 
\frac {\cos \kappa - 1}{\kappa^{2}} \right ),
$$
$$
f_{2}(\kappa) = \sin \kappa + 2 \left(\frac {\cos \kappa}{\kappa} -
\frac {\sin \kappa}{\kappa^{2}} \right ),
$$
and $c_{1,2} (k)$ are the amplitudes of the  growing and decaying modes
respectively . The sound velocity here is $\beta = 1/ \sqrt{3} \simeq
1$, so, $\kappa$ function presents a ratio of the sound pass during the
cosmological time $t$ to the perturbation scale $l_{k}$ which is about
the horizon to perturbation scale ratio.

The perturbations of the gravitational potential or metric
perturbations are as follows (exact definitions are given below):
$$
h_{1} = c_{1} \frac {1 - \cos \kappa}{\kappa^{2}} \;\;\;\; {\rm for \;\;
the \;\; growing \;\;  mode},
$$
$$
h_{2} = c_{2} \frac {\sin \kappa}{\kappa^{2}} \;\;\;\; {\rm for \;\; the \;\;
decaying \;\; mode} \eqno(9)
$$

An important feature of these perturbations is the following. Neither
growing nor decaying modes increase catastrophically in time. Both of
them are described by $\sin$ and $\cos$ functions, so, if $c_{1}$ and
$c_{2}$ are less than unity (and it should be this way, otherwise
$h_{1,2}$ would be large at smaller time) then both these modes are
just sound waves with constant in time amplitudes $c_{1}$ and $c_{2}$
and with different time phases.

First of all, this result means that the HFU is absolutely gravitationally
stable against small perturbations: if initial perturbations are less than
unity then they remain small forever till equation $p = \epsilon/3$ holds.

A more elegant proof of this important conclusion may be done with help
of the $q$-scalar (Lukash, 1980) which is, generally, gauge invariant
combination of matter velocity and gravitational perturbation
potentials. Below, we shall see that potential perturbations in FU are
totally described  by this scalar (and back: all matter and metric
perturbations can be expressed as functions of $q$). The physical
meaning of the $q$-scalar easily follows from its definition: for large
scale, $l_{k} > l_{H}, \;\; q$ is mainly the gravitational potential (matter
effects are not important), while inside horizon, $l_{k} < l_{H}$,
gravitational perturbations are negligible and $q$ is just the matter
(velocity) potential.

In the HFU the $q$-field obeys the following equation:
$$
\ddot{q} + 3H \dot{q} - \frac {1}{3a^{2}} \Delta q = 0, \eqno(10)
$$
where $H = \dot{a}/a$ is the Hubble function and $\Delta=\partial^{2}/ 
\partial \vec{x}^{2}$ is the spatial Laplacian. Transformations $\bar{q} =
aq,\;\; (')=d/d\eta=ad/dt$ reduce eq. (10) to (note, that $a\sim\eta$ for HFU)
$$
\bar{q}^{''} - \frac {1}{3} \Delta \bar{q} = 0, \eqno(11)
$$
which is just the non-gravitating acoustic wave equation in the flat
spacetime $(\eta, x)$:
$$
\bar{q}_{k} \sim c_{1} \sin \kappa + c_{2} \cos \kappa. \eqno(12)
$$

Eqs. (1,10,11,12) indicate the conformal invariance of potential
perturbations in HFU and, as a result, the conservation of the
adiabatic invariant --- the total number of phonons, the sound wave
quanta --- which proves the stability of the HFU expansion against small
matter perturbations \footnote {We do not go into further detail about
this stability effect since we have emphasized it many times in our
previous lectures. Mention only that the increase in time of the
density contrast at $\kappa < 1$ (see eq. (8)) which some people interpret
as an instability period, simply corresponds to the period of time of
the monotonic change of the oscillatory function. (None of the
sin-oscillations manage for $\kappa < 1$). Thus, to speak on the instability
in this case is as incorrect as to speak on the instability of a
mathematical pendulum when it moves, say, out of its stable point for
the time which is less than the oscillatory period. Returning to our
case, note that the potential energy of such a pendulum at $\kappa < 1$ is
all in the gravity (see eqs. (9)). The Jeans thinking fails here
because  the $\kappa < 1$ region is purely relativistic one.}. Note in this
connection that gravitational waves have a similar invariance property 
(see Grichshuk 1974)  but we do not discuss them here.

The lumps of the matter in these sound waves start growing only after
the equality epoch $(l_{H} \sim 10^{4} yrs)$ when the non-relativistic
particles become to dominate in the expansion and the pressure falls
down in comparison with the total density. This process develops due to
the Jeans gravitational instability causing the fragmentation of the
medium into separate bodies at the late stages of the expansion $(l_{H}
\sim 10^{9} - 10^{10} yrs)$.

We shall not discuss here these late processes of galaxy formation. For
us the following is important: for the formation of large structures
(superclusters and clusters), we need a definite  amplitude of the sound 
waves $\sim 10^{-4} - 10^{-5}$ in the linear scales which encompass the 
number of baryons big enough for these structures formation. So, $c_{1}$ 
or/and $c_{2}$ must be of the order of $10^{-4}$ on these scales.

It is a very serious demand on the initial perturbations. Indeed, when
$t$ is small, $\kappa \ll 1$, we have (see eqs. (9,12)):
$$
c_{1} \ll 1,
$$
$$
c_{2} \ll \kappa. \eqno(13)
$$

From these expressions we can see that $c_{2}$ must be extremely small
and cannot be of the order of $10^{-4}$. So, we need in fact the following
equations to be met for $\kappa \ll 1$:
$$
\hspace*{-5 mm} (i) \,\, c_{1} \gg c_{2},
$$
$$
(ii) \,\, c_{1} \sim 10^{-4}. \eqno(14)
$$
But both of them look very strange.

Indeed, any general natural initial conditions assume a random time phase 
state for the seed fluctuations
$$
c_{1} = c_{2} \ll \; 1. \eqno(15)
$$
E.g., the first eq. in (15) holds for vacuum or thermal fluctuations.
More of that, any natural fluctuations in hot gravitating medium imply
that $c_{1}$ and $c_{2}$ are dozers orders of magnitude less than $10^{-4}$.

The last point is demonstrated with help of the following example. Let
us suppose that the origin time of the fluctuations is the Planckian
one and let us denote $k = 1$ for $l_{pl}$. Then on the galactic scale
$k_{gal} \sim 10^{-26}$. Now, let us take a  thermal fluctuation
spectrum at this moment with the Planckian temperature and, thus, the
maximum at $l \sim l_{pl}$. Then the amplitude of the perturbations for
$k < 1$ would be proportional to $k^{3/2}$ and, on the galactic scale,
it would be $\sim 10^{-40}$. So, $c_{1}$ has to be $\sim 10^{-40}$ and
it is 35 orders of magnitude less than we need.

Our results are the following:
\begin{itemize}
\item[1)]
The classical cosmology of the hot VEU has principal difficulties in
the explanation of origin of the PCPs. Both requirements provided by
eqs. (14) for the large scale structure formation, cannot be naturally
explained within the frameworks of the HFU.
\item[2)] To account for the appearance of the  PCPs at the hot Universe
expansion period we need, as a necessary condition, to reject the $p =
\epsilon/3$ equation of state at the VEU stage. The modern cosmology
provides for a variety of the possibilities of such type: from
quantum-gravity effects to vacuum phase transitions, cosmic strings,
textures, etc. Here we consider the most general conditions for the
parametric amplification effect appearing in theories with one scalar
field $\varphi$ coupled to gravity in the minimal way.
\end{itemize}

Parametric amplification means the production of the gravitating
potential inhomogeneities (PCPs) in a non-stationary gravitational
background of the expanding Universe: large scale dynamic gravitational
field parametrically creates (amplifies) the small scale perturbation
fields. Mathematically, potential perturbations of FU with a general
expansion law are governed by the $q$-scalar which, after the conformal
transformations $(q \rightarrow \bar{q}, \, t \rightarrow \eta)$, meets
the following equation:
$$ 
\Box_{\beta} \bar{q} = U \bar{q}, \eqno(16)
$$
where $\Box_{\beta} = \partial^{2}/ \partial \eta^{2} - \beta^{2}\Delta$ 
is the light $(\beta = 1)$ or sound type $(\beta < 1)$ d'Alambertian 
operator in the conformal spacetime and $U = U(\eta)$ is the effective 
potential of $q$-field, which is a function of the expansion rate of the FU. 
Eq. (16) is a type of the parametric equation in mathematical analysis 
capable to amplify the fields with scales $k\, {}^{<}_{\sim}\, U^{1/2}$ 
which are usually outside or about the horizon size (i.e., in the purely 
relativistic region).

Say, for the massless scalar field $\varphi$ with minimal coupling the
effective potential is $U = a^{''}/a$, thus, the typical frequency is just
the horizon one (see eq. (16)):
$$
\frac{U^{1/2}}{a} = \frac{(a^{2}H \dot{)}^{1/2}}{a} = H\left(2 -
\frac{dl_H}{dt}\right)^{1/2} \sim H. \eqno(17)
$$

For the HFU, $(a\sim\eta)$, the effective potential is identically zero 
$U=0$, which reduces eq. (16) to eqs. (11,12) considered before. In
this case we can define the vacuum state of the $q$-field for all
spatial frequencies and introduce, for instance, a standard technics for
the scattering problem with $\mid in \rangle$ and $\mid out \rangle$,
vacua, and so on, to see how many phonons are spontaneously created
during expansion, which are their spectrum, etc.

Further applications depend on the sign of the second derivative of the
initial scale factor.

The point is that this sign can give us the idea about which scale expands
faster: the perturbation scale $l_{k}$ or the horizon $l_H$ (see eqs.
(4,5)). Indeed, the first derivative of their ratio is just proportional 
to the second derivative of the scale factor:
$$
\left( \frac{l_k}{l_H} \right)^{.} \sim \ddot a. \eqno(18)
$$

So, if $\ddot{a} < 0$ at the beginning, then the galactic scales are
found initially outside the horizon, and the Cauchy initial data should
be set up outside the horizon as well. On the contrary, if $\ddot{a} >
0$, then the initial conditions for the scales of interest can be set
up inside the horizon.

We shall investigate both cases. The qualitative result is as follows:
under natural initial conditions met by eq. (15) (e.g., $\mid in \rangle$
vacuum state for the $q$-field) it is the growing mode of perturbations
that is finally created in the $\mid out \rangle$ state due to the parametric
effect. So, the resulting perturbation field is described  by the first
line of eq.(14) with the $c_{1}(k)$ spectrum depending on the expansion
factor behaviour at time period when the parametric amplification
condition was met $(k \le U^{1/2}\sim aH)$.

The Lagrangian theory and the quantization of  potential perturbations
in FU are considered below. The next chapter deals with some cosmological
applications.

\subsection{Lagrangian Theory and Perturbations}
Let us consider a scalar field $\varphi = \varphi (x^{i})$ with the
Lagrangian density depending on $\varphi$ and its first derivatives in
the following general form:
$$
L=L(w,\varphi),\;\;\; w^2=\varphi_{,i}\varphi^{,i}=\varphi_{,i} 
\varphi_{,k} g^{ik} , \eqno(19)
$$

The action of the gravitating $\varphi$ field is as follows:
$$
W[\varphi, g^{ik}]= \int (L -\frac{1}{2} R) \sqrt{-g} \; \; d^{4} x,
\eqno(20)
$$
where $g_{ik}$ and $R_{ik}$ are the metric and Ricci tensors
respectively, $R = R^{i}_{i}, \, g = det(g_{ik})$. Variations of eq.
(20) over $\varphi$ and $g^{ik}$ in extremum give the clasical
equations of motions of the $\varphi$ field
$$
\left( \frac{n}{w} \varphi^{,i}\right)_{;i} + n \nu = 0, \eqno(21)
$$
and of the gravitational field created by the $\varphi$-field-source
$$
G_{ik} = T_{ik}, \; \; \; T_{ik} = \frac{n}{w} \varphi_{,i}
\varphi_{,k} - g_{ik} L, \eqno(22)
$$
where $n=\partial L/\partial w$, $n\nu=-\partial L/ \partial \varphi$,
$G_{ik} = R_{ik} - Rg_{ik}/2$ and (;) is the covariant derivative in
metric $g_{ik}$. Note, that eq. (21) can be obtained from the Bianchi
identities $T^{k}_{i;k} = 0$, as well.

Useful constructions are the comoving (to $\varphi$-field) energy
density and the total pressure of the $\varphi$ field
$$
\epsilon = T_{ik} \varphi^{,i} \varphi^{,k}/ w^{2} =n w - L ,
$$
$$
p = \frac{1}{3} (\epsilon - T) = L , \eqno(23)
$$
where $T = T^{i}_{i}$. Also, the following equations are valued
$$
\epsilon + p = n w ,\;\;\;\; \beta^{-2} = \frac{w}{n}
\frac{\partial^{2} L}{\partial w^{2}} = \frac{\partial \ln n}{\partial\ln w},
$$
$$
m^{2} = - \frac{w}{n} \; \frac{\partial^{2} L}{\partial \varphi^{2}},
\;\;\; \Gamma = \frac{w}{n} \frac{\partial^{2} L}{\partial w \partial
\varphi} = w \frac{\partial \ln n}{\partial \varphi}. \eqno(24)
$$
(Functions $m^{2}$ and $\beta^{2}$ can be negative).

When considering linear perturbation theory $\varphi$ and $g^{ik}$ are
presented as sums of some known functions (the background $({}^{o})$
solution) and small perturbations $\phi$ and $h_{ik}$:
$$
\varphi = \varphi^{(o)} + \phi, \;\;\; g^{ik} = g^{ik(o)} - h^{ik}.
\eqno(25) 
$$

Below, we consider the classical backgrounds (eqs. (21,22) are met
automatically in $({}^{o})$ order) and the perturbations can be quantum
ones.

The Lagrangian of perturbation field is  got by expanding the integrand
of eq. (20) up to the second order in $\phi$ and $h^{ik}$ with the
total divergent terms excluded (see Appendix A):
$$
W^{(2)} [\phi, h^{k}_{i}] = \int L^{(2)} \sqrt{-g^{o}} \; d^{4}x,
$$
$$
L^{(2)} = L^{(2)} (v, \psi^{k}_{i}) = \frac{\epsilon + p}{2} [v_{i}
v^{i} + \chi^{2}(\beta^{-2} - 1) - 2 v_{i} \psi^{i}_{k} u^{k} + v(\nu
\psi - m^{2}v + 2 \Gamma \chi)] +
$$
$$
+ \frac{\epsilon - p}{2} (\psi_{ik} \psi^{ik} -
\frac{1}{2} \psi^{2}) + \frac{1}{8} (\psi_{ik;l} \psi^{ik;l} - 2
\psi_{ik;l} \psi^{il;k} - \frac{1}{2} \psi_{,l} \psi^{,l}), \eqno(26)
$$
where $v = \phi/w , \, v_{i} = v_{,i} + v(w_{,i}/w)$, 
$u_{i} = \varphi^{(o)}_{,i}/w^{(o)}$, 
$\chi = v_{i} u^{i} - h_{ik} u^{i} u^{k}/2 = \delta w /w$, 
$\psi^{k}_{i} =h^{k}_{i} - h \delta^{k}_{i}/2$, 
$h = h^{i}_{i} = - \psi$. (Hereafter all manipulations with indeces are 
carried out with help of the background metric tensors $g^{(o)}_{ik}$ and 
$g^{ik(o)}$, and background index $({}^{o})$ is omitted where possible). 
Obviously,
$$
\frac{\delta p}{\epsilon + p}  = \chi - \nu v, \;\; \frac{\delta
\epsilon}{\epsilon + \rho} = \beta^{-2} \chi + (\nu + \Gamma)v. \eqno(27)
$$

The clasical field equations which couple the metric and scalar
perturbations, can be obtained either when the first variations of the
action (26) are taken equal to zero or, directly, while expanding eqs.
(21,22) to the linear order terms. Generally, these equations describe
three oscillators coupled to each other through the background shear
and vorticity (6-order in time equation system): one oscillator is the
scalar potential perturbations and the other two are just two
polarizations of the gravitational waves. \footnote {The vortex
perturbations (if any) are standardly found as the first integrals of
the 6-order equation system}.

To find the physical degrees of freedom of the perturbation fields and
to approach the problem of the PCP origin, the following steps have to
be developed.
\begin{itemize}
\item[(i)] 
Gauge invariant functions must be introduced instead of $\phi$ and $h_{ik}$

The point is that, although the original fields, scalar $\varphi$ and
tensor $g_{ik}$, are genuine by definition, their decomposition into
background and perturbation parts is not unambiguous at all. Indeed, if
we transform infinitesimally the reference system,
$$
x^{i} = \tilde {x}^{i} + \xi^{i}, \eqno(28)
$$
where $\xi^{i} = \xi^{i}(x^{k})$ are small arbitrary functions, then
the new separation in the coordinates $\tilde {x}^{i}$ will take the
following form:
$$
\varphi = \varphi^{(o)}(x^{i}) + \phi = \varphi^{(o)}(\tilde{x}^{i}) +
\tilde{\phi}, \; \; \; \tilde{\phi} = \phi + w \xi_{i} u^{i},
$$
$$
g_{ik} dx^{i} dx^{k} = \tilde {g}_{ik} d \tilde{x}^{i} d
\tilde{x}^{k}, \; \; \; \tilde{h}_{ik} = h_{ik} + \xi_{i;k} +
\xi_{k;i}, \eqno(29)
$$
where the background metric $g^{(o)}_{ik}$ has the same functional
dependence in the new coordinates $g^{(o)}_{ik}(\tilde{x}^{l})$ as that
in the old ones $g^{(o)}_{ik}(x^{l})$.

To develop the gauge invariant theory one has, first, to expend the
perturbation tensor $h_{ik}$ over the irreducible representations of
the background geometry to mark off the scalar and gravitational wave
polarizations and, second, to find the appropriate gauge invariant
(i.e., independent of the transformations (29)) linear superpositions
of the perturbation functions.

\item[(ii)] The Lagrangian and Hamiltonian formalisms of the perturbation
fields should be developed on the basis of the gauge invariant theory.

An important point here is to obtain the canonical field variables
accounting for the physical degrees of freedom.

\item[(iii)] Secondary quantization of the perturbations and cosmological
applications can be considered in connection of the PCP problem.

Here, we are going to analyze all these points for FU background
metrics. In this case
$$
g_{oi} = u_{i} = \delta^{o}_{i}, \;\;\;\;  g_{\alpha \beta} = -a^{2}
\gamma_{\alpha \beta},
$$
$$
H^{2} = \frac{1}{3} \epsilon - \frac{K}{a^{2}}, \;\;\;\; \dot H =
-\frac{\epsilon + p}{2} + \frac{K}{a^{2}},
$$
$$
\frac {\dot{n}}{n} + 3H + \nu = 0, \;\;\;\; v_{i} = v_{,i} -\beta^{2}
v(3H + \nu) u_{i}, \eqno(30)
$$
$$
\chi = \dot{v} - \beta^{2} v(3H + \nu+ \Gamma), \;\;\;\; \frac{\delta
\epsilon}{\epsilon + p} = \beta^{-2} \dot{v} - 3Hv,
$$
where $\gamma_{\alpha \beta} = \gamma_{\alpha \beta} (x^{\gamma})$ is
the metric tensor of the homogeneous isotropic 3-space with the spatial
curvature $K = 0, \pm 1$ (manipulations with the Greek indices are done
with help of $\gamma_{\alpha \beta}$). All the perturbation types
evolve independently of each other in the linear approximation since
the background shear and vorticity are identically zero. Below, only
potential perturbations are considered.

To avoid formal mathematical constructions we try to use here spatially
flat FU ($\gamma_{\alpha \beta} = \delta_{\alpha \beta}, \; K = 0,$ see
eq. (1)) and synchronous reference system for its perturbations
($h_{oi} = 0$) if no other cases are pointed out explicitly. The
general necessary formulae are given in Appendix B.
\end{itemize}

\subsection{Potential Perturbations in Friedmann Cosmology}
Let us obtain equations of motion of the gauge invariant potential
perturbations directly from the linearized Einstein eqs. (22), and
their Lagrangian from eqs. (26).

General metric perturbations in the synchronous reference system are
presented in terms of two gravitational potential, $A$ and $B$:
$$
ds^{2} = dt^{2} - a^{2} (\delta_{\alpha \beta} + h_{\alpha \beta})
dx^{\alpha} dx^{\beta},
$$
$$
h_{\alpha \beta} = A \delta_{\alpha \beta} + B_{, \alpha \beta}. \eqno(31)
$$

Since this metric is governed by the scalar field (19), we have three
perturbation potentials $v, A$ and $B$ entering the field equations. In
fact, only one of them is independent.

We will consider as the independent one a gauge invariant scalar $q$
which is a linear combination of the perturbation potentials. Let us
first define the $q$-scalar and then, using the low order equations,
relate inversely this scalar to $v, A$ and $B$. The gauge freedom in
the choice  of these potentials follows from eqs. (29):
$$
\tilde{v} = v + \frac{F}{2}, \;\; \tilde{A} = A + HF, \;\;\; \tilde{B}
= B + F \int \frac{dt}{a^{2}} + G, \eqno(32)
$$
where $F$ and $G$ are small arbitrary functions of the space
coordinates, and
$$
2\xi_{i} = F u_{i} - a^{2} (F_{, k} \int \frac{dt}{a^{2}} + G_{, k})
P^{k}_{i}, \eqno(33)
$$
is the most general form of the $\xi_{i}$-vector in the synchronous
gauge (Landau $\&$ Lifshitz 1967), $P^{k}_{i} = \delta^{k}_{i} - u_{i}
u^{k}$ is the projection tensor.

Eqs. (32) show that the following function is gauge invariant:
$$
q = A - 2 H v, \eqno(34)
$$
In fact, this function is independent of any other gauge as well (see 
Appendix B) which proves that $q = q(x^{i})$ is a 4-scalar in 
the unperturbed FU.

To derive the inverse transformations and the equation of motion for
$q$-field we will need only the low-order (in time) Einstein equations:
$$
\delta G^{o}_{o} = H \dot{h} - \frac{\Delta A}{a^{2}} = \delta
\epsilon, \eqno(35a)
$$
$$
G_{o \alpha} = - \dot{A}_{, \alpha} = (\epsilon + p) v_{, \alpha},
\eqno(35b)
$$
$$
\delta G^{\beta}_{\alpha} - \frac{1}{3} \delta^\beta_\alpha 
\delta G^{\gamma}_{\gamma} = \frac{1}{2a^{2}} (C^{, \beta}_{, \alpha} -
\frac{1}{3} \Delta C \delta^{\beta}_{\alpha}) = 0, \eqno(35c)
$$
where $h = 3A + \Delta B$ and $C = A - (\dot{B} a^{3} \dot{)}/a$. The
first two eqs. (35a,b) are just the conservations of energy and
momentum, respectively, whereas the last one (35c) states the
Pascalian condition (the absence of pressure anisotropies for
$\varphi$-field). In the class of functions under interest eq. (35c) have 
only the trivial solution\footnote {The formal solution is $C = f
\vec{x}^{2} + \vec{g} \vec{x} + e$, where $f, \vec{g}$
and $e$ are functions of time. The last two terms in the r.h.s. can be
excluded because they do not enter the original $h_{\alpha \beta}$
functions. The quadratic term can be ascribed only to $B$ potential
which results in $h_{\alpha \beta} \sim \tilde{f} (t) \delta_{\alpha
\beta}$, the latter excluded by redefining the scale factor.}:
$$
C = 0 \eqno(36)
$$
which relates $A$ and $B$ potentials straightforwardly.

Making use of eq. (35b), we can now express functions $v,
A$ and $B$ in terms of the $q$-scalar:
$$
v = \frac{1}{2}(Q - \frac{q}{H}), \;\;\;\; A = HQ, \;\;\;\; \dot{B} =
a^{-2} Q - a^{-3} P,
$$
$$
Q = \int \gamma q dt, \;\;\; P = \int a \gamma q dt,
$$
$$ 
\gamma = - \frac{\dot{H}}{H^{2}} = \frac{\epsilon + p}{2H^{2}} =
\frac{3}{2} \frac{\epsilon + p}{\epsilon} . \eqno(37)
$$

Functions $Q = Q(x^{i})$ and $P = P(x^{i})$ are the $q$-integrals over
the Friedmannian world line $dt = u_{i}dx^{i}$. They are determined up
to the accuracy of some additive functions of the space coordinates.
This freedom for the $Q$-function is just the gauge one (see (32)). The
$P$ scalar is gauge invariant, so its "ambiguity" is physically
meaningfull and related to a certain class of perturbations of the
Bianchi type I model. To prove it we need another relation between $q$
and $P$ functions which we are going to get from eq. (35a).

Let us recover the energy perturbation using eqs. (30, 37):
$$
\frac{\delta\epsilon}{\epsilon+p}=-\frac{\dot q}{2H\beta^2}-3Hv. \eqno(38)
$$

Now, substituting it into eq.(35a), we have the key equation for $q$-scalar: 
$$
\gamma a^3 \dot{q}=\beta^2\Delta P   \eqno(39)
$$
which is obviously the GR-analog of Poison equation.

Let us first assume that $\beta \ne 0$. Comparison of eqs. (37, 39)
shows that $P(t, \vec{x})$ is specified by the $q$-scalar up to accuracy of
additive harmonic function of spatial coordinates $P(\vec{x})$:
$$
\Delta P(\vec{x}) = 0. \eqno(40)
$$

In the class of the uniformly limited (in 3-space) functions $h_{\alpha
\beta} \sim P(\vec{x})_{, \alpha \beta}$ the solution is a bilinear form with
zero trace:
$$
P(\vec{x}) = a_{\alpha \beta} x^{\alpha} x^{\beta}, \;\;\;\; a_{\alpha \beta}
= {\rm const}, \;\;\;\; a^{\alpha}_{\alpha} = 0. \eqno(41)
$$

Thus, potentials $v, A$ and $B$ are reconstructed from the given
$q$-scalar but a partial solution that does not vanish under gauge
transformations:
$$
v = A = 0, \;\;\; B = P(\vec{x}) \int \frac{dt}{a^{3}}. \eqno(42)
$$

Appendix C demonstrates that these perturbations (42) are homogeneous
and belong to the Bianchi type I cosmological model.

For $\beta = 0$, function $P(\vec{x})$ is arbitrary and eqs. (42) describe
the decaying mode of perturbations. The growing mode is determined by
another arbitrary function of the space coordinates, $q(\vec{x})$. The
general solution in this case is
$$
v = 0, \;\;\;\; Q = \frac{q(\vec{x})}{H},
$$
$$
P = q(\vec{x}) \int a \gamma dt + P(\vec{x}), \;\;\;\; \delta \epsilon =
\frac{H \Delta P}{a^{3}}. \eqno(43)
$$

So, with all the above said we may conclude that, for $\beta \ne 0$,
the $q$-scalar is totally responsible for the evolution of the physical
potential perturbations in spatially flat Friedmann model. Division over
$\beta^{2}$ and differention of eq. (39) (with eq. (37) for the $P$-scalar
taking into account) gives the second-order equation for the $q$-scalar:
$$
\ddot{q} + \left(3H + 2 \frac {\dot{\alpha}}{\alpha}\right) \dot{q} -
\left(\frac{\beta}{a}\right)^{2} \Delta q = 0, \eqno(44)
$$
where $\alpha^{2} = \gamma/2 \beta^{2}$.

Now, we can derive the Lagrangian density for the $q$-field.
Substituting eqs. (37) into eq. (26) and leaving out full divergent
terms, we have after rather lengthy calculations:
$$
W^{(2)} [\phi, h^{k}_{i}] = W[q] = \int L(q) a^{3} dt d^{3}x,
$$
$$L(q) = \frac{1}{2} \alpha^{2} \left(\dot{q}^{2} -\left(\frac{\beta}{a}
\right)^{2}q_{, \alpha} q^{, \alpha}\right), \eqno(45)
$$
where $L(q)$ and $L^{(2)}$ differ each from the other only in the
divergent terms.

Eqs. (45) evidence that $q$-scalar is the unique single canonical
variable for physical degree of freedom of potential perturbations in
the FU driven by scalar field of type (19). The Lagrangian density depends
only on the first derivatives of $q$-field. However, if $\alpha \ne
{\rm const}$ then $q$ acquires a mass. Endeed, introducing the following 
transformation 
$$
\tilde{q} = \alpha q, \eqno(46)
$$
we may rewrite the Lagrangian in the form of standard scalar field with
square-mass $\mu^{2} = - \ddot{\alpha}/ \alpha$:
$$
L(q)=\frac{1}{2}\left(\dot{\tilde{q}}-\left(\frac{\beta}{a}\right)^{2}
\tilde{q}_{,\alpha}\tilde{q}^{,\alpha}-\mu^{2}\tilde{q}^{2}\right)\eqno(47) 
$$

Appendix B confirms eqs. (44,45) for general case. The covariant 
generalization is as follows:
$$
(D^{ik} q_{,i})_{; k} = 0 \;\;\; D_{ik} = \frac{1}{2} \alpha^{2}
(u_{i} u_{k} + \beta^{2} P_{ik}), \eqno(48)
$$
$$
L(q) = \frac{1}{2} D^{ik} q_{, i} q_{, k}, \;\;\; W[q] = \int L(q)
\sqrt{-g} \; d^{4}x, \eqno(49)
$$
where $\alpha = (\epsilon + p)^{1/2}/2 \beta H$, $P_{ik} = g_{ik} -
u_{i} u_{k}$, $g = {\rm det}(g_{ik})$, all Friedmann functions. 

Before going to the next point, we give some relations for density
perturbations in different systems most frequently used in literature. 

For the synchronous gauge we have from eq. (38):
$$
\frac{\delta \epsilon}{\epsilon + p} = - \frac{\dot{q}}{2H \beta^{2}}
+ \frac{3}{2}(q - HQ). \eqno(50)
$$

For the comoving reference system with the synchronized time $(\tilde
v_{i} \sim u_{i})$:
$$
\frac{\delta \tilde \epsilon}{\epsilon + p} = - \frac{\dot{q}}{2H
\beta^{2}} , \eqno(51)
$$
where
$$
d \tilde s^{2} = \left(1 + \frac{\dot{q}}{H}\right)d \tilde t^{2} -a^{2}
(\delta_{\alpha \beta} (1 + q) + \tilde B_{, \alpha \beta})d
\tilde x^{\alpha} d \tilde x^{\beta},
$$
$$
\tilde t = t + v, \;\;\;\ \tilde x^{\alpha} = x^{\alpha} + \int
\frac{v^{, \alpha}}{a^{2}} dt, \eqno(52)
$$
$$ 
{\dot {\tilde B}} = \frac{q}{Ha^{2}} - \frac{P}{a^{3}}.
$$

For the Newtonian gauge (zero shear reference system, $\tilde{\tilde{B}}=0$):
$$
\frac{\delta \tilde {\tilde{ \epsilon}}}{\epsilon +p} = - \frac{\dot{q}}{2H
\beta^{2}} + \frac{3}{2}(q - \frac{HP}a), \eqno(53)
$$
where
$$
d \tilde {\tilde {s}}^{2} = 
\left(1 - \frac{HP}a\right)d \tilde {\tilde{t}}^2 - a^{2} \left(1 +
\frac{HP}a\right) \delta_{\alpha \beta} d \tilde {\tilde{ x}}^{\alpha} 
d \tilde {\tilde {x}} ^{\beta}
$$
$$
\tilde{\tilde{t}} = t + \frac{1}{2} a^{2} \dot{B}, \;\;\ \tilde {\tilde
{x}}^{\alpha} + \frac{1}{2} B^{, \alpha}.
$$

For scales in the horizon $(k \eta \gg 1)$ all the three expressions
for $\delta \epsilon$ coincide since the leading term is the first one 
$(\delta \epsilon \sim\dot{q})$. In the relativistic region $(k\eta\le1)$, 
$\delta\epsilon$ depends explicitly on spatial slice given.

\subsection{Quantization and Conformal Non-Invariance}
Taking in mind eqs. (48,49) one can formally consider the $q$-field as a 
test scalar field in the Friedmann models. It allows for a standard 
development of the Hamiltonian formalism.

First, we can construct the Hilbert space of all complex solutions of
eq. (48) with the scalar product
$$
(q_{1}, q_{2}) = \int \limits_{\Sigma} J^{i}_{12} d \Sigma_{i},
$$
$$
J^{i}_{12}=iD^{ik}(q^{\ast}_{1}q_{2,k}-q^{\ast}_{1,k}q_{2}),\eqno(54) 
$$
where $d \Sigma_{i}$ is the invariant measure on a Cauchy-hypersurface
$\Sigma$. The integral in eq. (54) does not depend on $\Sigma$ choice
because of the 4-flux conservation law $J^{i}_{12;i} = 0$.

Next, the canonically conjugate gauge invariant scalar is introduced:
$$ 
\sigma = \sigma (x^{i}) = \frac{\partial L(q)}{\partial \dot{q}} =
\alpha^{2} \dot{q}. \eqno(55)
$$

Further steps to the constructing the field Hamiltonian and canonical
quantization are as simple as that in the case of any other scalar
field. We would like to emphasize here two points.

The quantization is based on the simultaneous commutation relation for
the canonically operators $q$ and $\sigma$:
$$
[q(t,\vec{x}), \sigma (t, \vec{x}^{'})] = q \sigma - \sigma q = i \sqrt{-g} 
\;\; \delta(\vec{x} - \vec{x}^{'}). \eqno(56)
$$

This equation can be compared with the commutator between the velocity
potential and density perturbation operators of sound waves in the
nongravitating static matter, $[v, \delta \epsilon] = i
\delta(\vec{x} - \vec{x}^{'})$ (see Lifshitz $\&$ Pitaevski 1978). 

It is worth while rewriting eq. (48) in terms of the conformal
coordinates $(\eta, x)$ for the conformal field $\bar{q}$:
$$
\Box_{\beta} \bar{q} = U \bar{q}, \;\;\;\; \bar{q} = \alpha a q, \eqno(57)
$$
where $\Box_{\beta} = \frac {\partial^{2}}{\partial \eta^{2}}-\beta^{2} 
\Delta$ is the d'Alambertian operator in the Minkowski metric $d\bar{s}^{2} 
= ds^{2}/a^{2} = d \eta^{2} - d\vec{x}^{2}$. The function $U =
U(\eta) = (\alpha a)^{''}/(\alpha a)$ plays a role of the effective
potential of scalar perturbations in FU. It is an unambiguous function
of the background expansion or, more precise, of the scale factor and
its time derivatives up to the forth order.
Note, that the Lagrangian for eq.(57)
$$
\bar{L} (\bar{q}) = \frac{1}{2a^{4}} (\bar{q}^{' 2} - \beta^{2}
\bar{q}_{, \alpha} \bar{q}^{, \alpha} + U \bar{q}^{2}) \eqno(58)
$$
coincides with $L(q)$ up to full divergent term.

The total energy of potential perturbations in the Friedmannian space
$t = {\rm const}$ --- the field Hamiltonian --- can be also presented in terms
of the conformal field:
$$
\frac{\bar H}{a} = a^{3} \int E d^{3} \vec{x},
$$
$$
E = \frac{1}{2a^{4}} (\bar{q}^{' 2} + \beta^{2} \bar{q}_{, \alpha}
\bar{q}^{, \alpha} - U \bar{q}^{2}), \eqno(59)
$$
where $E = E(\eta, \vec{x})$ is the local energy density of the $q$-field.
Note, that for the non-gravitating matter (or for short wavelengths $k
\eta \gg 1$) $E$ is analogous to the sound wave energy density:
$$
E \simeq \frac{(\epsilon + p)\vec{v}^{2}}{2} + \frac{(\beta \delta
\epsilon)^{2}}{2(\epsilon + p)}, 
$$
where $\vec{v} = (v^{, \alpha}/a)$ is the matter velocity.

Since this formal analogy with sound waves and the fact that PCPs,
which are just the resulting (after amplification) $q$-field, are usually
found at the beginning of the HFU expansion stage, we will call below the
$q$-field quanta as phonons. These cosmological phonons remind the standard
physics phonons only when phonon wavelength is inside the horizon $(k
\eta \gg 1)$ \footnote {Rigorously speaking, inside the sound
horizon, $k \eta \gg |\beta |^{-1}$ (see the d'Alambertian in
eq. (57)). For estimates, we assume in the main text that $\beta \sim
1$.}, in this case  the gravity is negligent and $q \sim v$ (see eq.
(34)). For large scales $(k \eta \le 1)$, matter effects are not
important and $q$ is mainly the gravitational field potential.

When scale factor is proportional to the conformal time, $q$-scalar
appears conformally coupled to FU (see eq. (57)):
$$
a \sim \eta/ \alpha: \;\; U(\eta) = 0. \eqno(60)
$$

In all the other cases $U \ne 0$ and the $q$-field is conformally
non-invariant. It means that $q$ interacts with background
non-stationary metric, which provides for the spontaneous and induced
production of phonons in the process of cosmological expansion. 

The secondary quantization of $q$-scalar results in the following expansion:
$$
q = \int d^{3} \vec{k}(a_{\vec{k}} q_{\vec{k}} + a^{\dag}_{\vec{k}} 
q^{\ast}_{\vec{k}}), \eqno(61)
$$
where
$$
(q_{\vec{k}}, q_{\vec{k}^{'}}) = [a_{\vec{k}}, a^{\dag}_{\vec{k}^{'}}] =
\delta(\vec{k} - \vec{k}^{'}),
$$
\hspace*{5cm} 
$$
(q_{\vec{k}}, q^{\ast}_{\vec{k}^{'}}) = [a_{\vec{k}}, a_{\vec{k}^{'}}] = 0, 
$$
$a_{\vec{k}}$ and $a^{\dag}_{\vec{k}}$ are the annihilation and creation
operators respectively,
$$
q_{\vec{k}} = q_{\vec{k}}(x^{i}) = \frac{\nu_{k}}{(2 \pi)^{3/2}
\alpha a} \; e^{i \vec{k} \vec{x}}, 
$$
and $\nu = \nu_{k}(\eta)$ satisfies the following equations
$$
\nu^{''}_{k} + (\beta^{2} k^{2} - U)\nu_{k} = 0, \;\;\;\; \nu_{k}
\nu^{\ast '}_{k} - \nu^{\ast}_{k} v^{'}_{k} = i.
$$

Below, we apply the theory of $q$-field for VEU. We shall assume $\mid in
>$ vacuum initial state for the $q$-field. To define it explicity, we will
consider in the next Chapter two cases for $\eta \rightarrow 0: \;
\ddot{a} < 0$ and $\ddot{a} > 0 $.

\section{Origin of Primordial Cosmological Perturbations}
Here, we consider some cosmological applications of the theory of $q$-field:
the scatterring problem and the problrm of the generation of PCPs in chaotic
inflation.

Let us separate explicitely the kinetic and potential terms in the Lagrangian:
$$
L = p(w) - V(\varphi). \eqno(62)
$$

From eqs. (23), we have
$$
6a\ddot{a} = -(\epsilon + 3p)= 2V(\varphi)-\left(\epsilon(w)+ 3p(w)\right).
$$ 
where $\epsilon (w) = nw - p(w)$. It is seen, that if $V(\varphi) > 0$
and $\epsilon(\omega) > 0$, then the case $\ddot{a} < 0$ can be realized 
only when the potential term is negligible, whereas for the apposite case 
$(\ddot{a} > 0)$ the potential term may play a dominant role. For this 
reason, we will consider two interesting for us asymptotics.

First, let us suppose that kinetic terms dominates the potential term
in general Lagrangian. The following {\it theorem} can be easily proved 
in this connection:

{\it The theory of a real scalar field with Lagrangian depending only on
the kinetic term,} 
$$
L = p(w), \;\;\;\; w^{2} = \varphi_{, i} \varphi^{,i}, \eqno(63)
$$
{\it is mathematically equivalent to the theory of potential motions of the
ideal fluid with arbitrary equation of state}
$$
p=p(w),\;\;\;\;\epsilon=\epsilon(w)=w\frac{dp(w)}{dw} - p(w). \eqno(64)
$$

{\it The 4-velocity of the ideal fluid is a time-like vector} $(w > 0)$:
$$
u^{i} = \frac{dx^{i}}{ds} = \varphi^{, i}/w . \eqno(65)
$$

{\it So, the $\varphi$-field acts here as the velocity potential}. 

In the other case $\ddot{a} > 0$, the potential term becomes important
and we may decompose $L$-function over small parameter $w^{2}$:
$$
L(w,\varphi)=- V(\varphi)+\frac{w^2}{2}W^2(\varphi)+0(w^{4}).\eqno(66a) 
$$

A simple redifinition of $\varphi$-field
$$
\varphi \Rightarrow \int W(\varphi) d \varphi
$$
reduces eq. (66a) to the following case with standard kinetic term :
$$
L = \frac{1}{2} \varphi_{, i} \varphi^{, i} - V(\varphi). \eqno(66b)
$$

Next Section deals with a general scattering approach for the $q$-field
(the first case can be solved only in this approximation). In the last
Section of this Chapter we consider the Lagrangian (66b) with
$V(\varphi) \gg |\varphi_{,i}\varphi^{,i}|$ initial condition.

\subsection{Scattering Problem for $q$-Field}
First, let us suppose that $V(\varphi) = 0$.

Applying the {\it theorem} to the flat FU, we see that for equation of state
$p = \epsilon/3$ the $q$-field is conformally coupled $(U = 0)$. Eqs.
(59) give the integral of motion
$$
a \sim \eta: \;\;\; \bar{H} = {\rm const}, \eqno(67)
$$
which means the conservation of the total number of phonons in the
process of the cosmological expansion.

In fact, the phonon numbers conserve at any frequency mode. Let us
dwell on it in a bit more detail.

At relativistic stage (67) phonons are presented by the following
choice of functions \footnote {We do not go into detail about such
standard for any quantized theory things as separation of the Hilbert
space in the positive and negative frequency subspaces, the Fock space
of states, Bogolubov transformations, etc.}:
$$
\nu_k=\frac 1{\sqrt{2\omega}} \exp(-i\omega \tau), \;\;\;\; 
\omega = \frac k{\sqrt 3},\eqno(68)
$$
where $a^{'} = {\rm const}, \;\; a/a^{'} = \tau = \eta + {\rm const}$, 
the prime $(')$ is the derivative in conformal time. The field Hamiltonian
is a sum over all quanta energies:
$$
H_{Reg} = \int d^{3} \vec{k} E_{k} N_{\vec{k}}, \eqno(69)
$$
where $E_{k} = \omega/a$ is the phonon energy and $N_{\vec{k}}=a^{\dag}_{
\vec{k}}a_{\vec{k}}$ is the operator of the number of phonons with physical 
momentum $\vec{k}/a$. The eigenvalues of the mean energy density operator
$$
E = \frac{\bar H}{a^{4}V}, \;\;\;\; V = \int d^{3} \vec{x} \eqno(70)
$$
are as follows
$$
E=\frac 1{(2\pi a)^3}\int d^3 \vec k E_k\left(n_{\vec k}+\frac 12\right),
\eqno(71)
$$
where $n_{\vec{k}}$ are the occupation numbers of the phonon states.

Eq.(67) also allows for introduction of the growing and decaying mode
operators: 
$$
C_1\equiv C_1(\vec k)=\sqrt{\frac{\omega}{2}} 
\left(\frac{a_{\vec{k}} -a^{\dag}_{- \vec{k}}}{ia{'}}\right), 
$$
$$
C_2 \equiv C_2(\vec k) = \sqrt{\frac{\omega}{2}} \left(
\frac{a_{\vec{k}} +a^{\dag}_{-\vec{k}}}{a^{'}}\right). \eqno(72)
$$

In terms of these operators the field expansions have the following
form (cf. eq. (12)):
$$
q = \frac {1}{(2 \pi)^{3/2}} \int d^{3} \vec{k} e^{i \vec{k} \vec{x}} 
\left(C_{1} \frac{\sin \omega\tau}{\omega \tau} + C_{2} \frac{\cos 
\omega \tau}{\omega\tau}\right), 
$$
$$
\bar H = \frac{1}{6} \epsilon a^{4} \int d^{3} \vec{k} (\mid C_{1} \mid^{2} +
\mid C_{2} \mid^{2}), \eqno(73)
$$
where $\mid C \mid^{2} = CC^{\dag} = C^{\dag} C$.

Now, let us calculate the number of phonons created at a period with some
arbitrary expansion law governed by the general Lagrangian (19):
$$
a = a(\eta), \;\;\;\; \eta_{1} < \eta < \eta_{2}. \eqno(74)
$$

We can do it directly taking into account the phonon numbers before
$(\eta \le \eta_{1})$ and after $(\eta \ge \eta_{2})$ this period and
just comparing them (the scattering problem).

As we have already seen, it is possible to calculate the occupation
numbers for any wavelength at a linear expansion stage (see eq. (67)).
Thus, to solve our problem, we should match the $a$-function (and its
first derivative $\dot{a}$ \footnote {We need this condition in order to
avoid the unwanted non-physical effects of creation in the matching
points. For further details about scattering problem, which are
standard for any test field theory, see Grib et al. (1980), Birrel $\&$
Davies (1981), and others.}) by the linear passes $a \sim  \eta$ at the
beginning $(\eta = \eta_{1})$ and at the end $(\eta = \eta_{2})$ of
the considered period.

So, in the resulting normalization (the Planck scale at Planck moment
of time is $k^{-1} = 1$) we have:
$$ 
a = \left\{\begin{array}{ll}
\tau = \eta = (2t)^{1/2},             & \eta \le \eta_{1}\\
a(\eta),                              & \eta_{1} < \eta < \eta_{2}\\
A_{1} \tau = A_{1} (\eta + \eta_{o}), & \eta > \eta_{2}, \; \eta_{o} =
{\rm const}
\end{array} \right. \eqno(75)
$$
where $\tau = \tau(\eta)$ is the conformal horizon (Hubble) time:
$$
\tau = \frac{a}{a^{'}} = \eta - \int_{\eta_{1}} \frac{aa^{''}}{a^{'2}}
d\eta=\frac{1}{2}\int\left(1 + \frac{3p}{\epsilon}\right) d \eta. \eqno(76)
$$

The value of the physical constant
$$
A_{1} = \frac{a^{'}(\eta = \eta_{2})}{a^{'}(\eta=\eta_{1})} \eqno(77)
$$
depends on the average expansion rate at period (74). Locally,
$A_{1}$-factor can be related to the conformal acceleration:  
$$
\frac{a^{''}}{a^{'}} = \lim \limits_{ {\textstyle
\eta_{\scriptscriptstyle 2}\rightarrow \eta_{\scriptscriptstyle 1}}}\left 
(\frac{A_{1} - 1}{\eta_2 - \eta_1}\right).\eqno(78)
$$

Let $a_{\vec{k}}$ and $b_{\vec{k}}$ be the phonon representations (68) 
diagalizing the Hamiltonian at stages $\eta \le \eta_{1}$ and $\eta \ge 
\eta_{2}$ respectively. Then
$$
b_{\vec{k}} = \alpha_{k} a_{\vec{k}} + \beta^{\ast}_{k} a^{\dag}_{-\vec{k}},
\;\;\;\; \mid \alpha_{k} \mid^{2} - \mid \beta_{k} \mid^{2} = 1, \eqno(79)
$$
where $\alpha_{k}$ and $\beta_{k}$ are Bogolubov coefficients. The
$\mid in \rangle $ and  $\mid out \rangle $ vacua are defined accordingly:
$$
a_{k} \mid in \rangle \; = 0,\;\; \;\; b_{k} \mid out\rangle\;\;=0.\eqno(80) 
$$

The Heisenberg state of the $q$-field is supposed to coincide with the
$\mid in \rangle \;\;$ vacuum. It means that there are no phonons (i.e.,
potential perturbations) initially. 

Taking average over the $\mid in \rangle \;\;$ vacuum state, we derive 
the mean occupation numbers of phonons spontaneously created at period
$\eta_{1} < \eta < \eta_{2}$,
$$
\langle b^{\dag}_{\vec{k}} b_{\vec{k}} \rangle = n_{k} \delta(\vec{k} - 
\vec{k}^{'}), \;\;\;\; n_{k} = \mid \beta_{k}\mid^{2}, \eqno(81)
$$
and ratio of the energy densities of the perturbations field to the
homogeneous cosmological field (see eqs. (70,71)):
$$
\frac{E_{Reg}}{\epsilon} = \frac 1{24 \pi^{3} A^{2}_{1}} 
\int d^{3} \vec{k} \;\;\omega \mid\beta_{k}\mid^{2}. \eqno(82)
$$

The factor $A^{-2}_{1}$ takes into account phonon energy cooling
during the expansion period (74).

Calculations of the produced spectrum is also straightforward:
$$
\langle q^{2} \rangle = \int^{\infty}_{o} \frac{dk}{k} q^{2}_{k}
$$
$$
q^{2}_{k} = \left(\frac{k}{\pi \tau A_{1}}\right)^{2} (\mid\beta_{k} 
\mid^{2} + Re (\alpha_{k} \beta^{\ast}_{k} e^{-2i \omega\tau})). \eqno(83)
$$

The spectrum dependence on the oscillatory exponent means that growing
and decaying modes are created non-equally (see eq. (73)). Now, we are
going to prove that under general condition $a^{''} > 0 \; (A_{1} > 1,$
see eq, (78)) which is most frequently met in the applications, it is
the growing mode of perturbations that is preferably created by the
parametric mechanism.

Indeed, our initial conditions generally impty $(\eta\le\eta_1,$ cf. eq.(15)):
$$
C^{2} = \langle \mid C_{1} \mid^{2} \rangle =  \langle \mid C_{2}
\mid^{2} \rangle \ll  1. \eqno(84)
$$

The situation is trivial in case of the acceleration $(\ddot{a} > 0)$
when the perturbation scales inflate from inside to outside the
horizon. The decaying mode, appearing originally at the horizon with
the same amplitude as the growing one, decays quickly for larger times
while the growing mode is frosen. It can be demonstrated with help of
general solution of the dynamic eq. (46) in large scales (the
Laplacian term is negligible):
$$
q=q_{1}(\vec{x}) + q_{2}(\vec{x}) \int \frac{dt}{\alpha^{2} a^{3}},\eqno(85)
$$
where $q_{1,2}(\vec{x})$ are arbitrary functions of the spatial
coordinates. The integral sharply converges to a constant in time function
$$
q = q(\vec{x}), \eqno(86)
$$
which describes the growing perturbation mode at $\eta\mid\nabla q\mid \ll
q$ 
for any expansion law.

Let us consider in a more detail the case when the initial conditions
(84) are set up outside the horizon. The general solution (85) helps
again. Comparing it with the $a$-representation functions $\nu_{k} =
\nu_{k}(\eta)$ which have the form (68) for $\eta \le \eta_{1}$, and
$$
\nu_{k} = \frac{1}{\sqrt{2\omega}} (\alpha_{k} e^{-i \omega \tau} +
\beta_{k} e^{i \omega \tau})
$$
for $\eta \ge \eta_{2}$, we obtain for the Bogolubov coefficients at $k
\tau_{1} \ll 1:$
$$
\alpha_{k} = \frac{1}{2} (A^{-1}_{1} + ig_{k} A_{1}), \;\;\;\; \beta_{k}
= \frac{1}{2}(A^{-1}_{1} - ig_{k} A_{1}), \eqno(87)
$$
where $g_{k} = \chi_{1} /k - i \;\;$ is the amplification coefficient,
$$
\frac{\chi_{1} \tau_{1}}{ \sqrt{3}} = 1 - \frac{\tau_{1}}{\tau_{2}A_1^2} -
\int^{\eta_{2}}_{\eta_{1}} (\alpha a)^{-2} d \eta = {\rm const}.
$$

Obviously, $g_{k} = \chi_{1}/k$ for $ g_{k} \gg 1$ and $g_{k} \sim
(\omega \tau_{1})^{-1}$ for $A_{1} \gg 1$. 

Substitution of eqs. (87) into eq. (83) gives the following spectrum
for $A_{1} \gg 1$:
$$
q_{k} = \frac{k^2}{2 \pi} \mid g_{k}\mid \frac{\sin \omega
\tau}{\omega \tau}. \eqno(88)
$$

The comparison with eqs. (73) reveals easily that only the
growing mode is created.

To clarify the physical meaning of this effect, let us relate directly
the $C_{1,2}$ operators in $a$ and $b$ representations (see eqs.
(72,79,87), $k \tau_{1} \ll 1)$:
$$
C^{(b)}_{1} = C^{(a)}_{1} + \frac{\chi_{1}}{k} C^{(a)}_{2}, \;\;\;\;
C^{(b)}_{2} = A^{-2}_{1} C^{(a)}_{2}. \eqno(89)
$$

So, if one begins with eq. (84) then, in the end, eqs. (89) give for
$A_{1} \gg 1$ and $g_{k} \gg 1$
$$
\langle \mid C_{1} \mid^{2} \rangle = g^{2}_{k} C^{2} \gg C^{2} \gg
\langle \mid C_{2} \mid^{2} \rangle, \eqno(90)
$$
what is just required by the galaxy formation theories (cf. eq.(14)).
In fact, the effect is a pure game of the mode mixing, it is not the
$q$-field itself that is created but rather the $q$-momentum (the time
derivative $\dot{\overline{q}})$. In other words one can say that the
parametric amplification effect brings about the creation of squeezed
state from initially random state.

So, as we could see, it is not a problem to produce the growing mode of
perturbations with  necessary amplitude. The amplification
coefficient is the larger the earlier HFU expansion is violated.
The typical spectra (88) go like
$$
q_{k} = Mk, \;\;\;\;\; k < M, \eqno(91)
$$
with the maximum amplitude corresponding to the mass scale $M \sim 
\tau^{-1}_{1}$ when the linear expansion law was broken for the first
time. (Note, that the spectrum (91) decreases to large wavelengths in
comparison with the Harrison-Zeldovich scale-free spectrum $q_{HZ} \sim
{\rm const} $). For $k \gg M$, the amplification coefficient is
exponentially small.

There are two points concerning eq. (91).
\begin{itemize}
\item[(i)] Initial (vacuum) conditions  are set up outside the horizon which
requires physical explanation.
\item[(ii)] To get the expansion factor required for typical scales $k \le M$
to be of the order of the galactic scales, it is necessary to ensure
the acceleration (inflationary) condition $\ddot{a} > 0$ at period (74).
\end{itemize}

In the latter case, the initial vacuum condition must be set up in the
adiabatic zone --- within the horizon --- which can be done independently
of the expansion law at the beginning.

\subsection{Generation of Perturbations on Inflation}
Let us consider Lagrangian (66) with a potential $V = V(\varphi) \in
C_{3}$ (at least, the first three derivatives determined). We shall
generally assume it to be a monotonically growing function of
$\varphi$ for $\varphi > \varphi_{o}$. Without loss of generality we
can put  $\varphi_{o} = 0$ and $V = dV/d \varphi = 0$ at $\varphi = 0$
which, with the symmetric condition $V(\varphi) = V(- \varphi)$, makes
a stable minimum of the potential $V(\varphi)$ at point $\varphi = 0$.
Under such normalization, our main assumption takes the following form:
$$
V > 0, \;\;\;\; \frac{dV}{d \varphi} > 0 \;\; {\rm for} \;\; 
\varphi > 0. \eqno(92)
$$

Also, we define three auxiliary functions of $\varphi$ related to the
potential derivatives:
$$
c = \frac{d \ln V}{d \ln \varphi}, \;\; e = \frac {d \ln c}{d \ln
\varphi}, \;\; f = \frac {d \ln e}{d \ln \varphi}.
$$

The simplest examples are $V_{1} = m^{2} \varphi^{2}/2$ and $V_{2} =
\lambda \varphi^{4}/4$ where constants $m, \lambda$ are the field mass
and dimensionless parameter, respectively. Evidently, for the power-law
potentials $V_{n} \sim \varphi^{2n}, \;\; c = 2n = {\rm const}$, and 
$e = f = 0$.

The background Friedmann quantities and eqs. (30) are
$$
\frac{2 \alpha^{2}}{3}=\left(1+\frac{2V}{\dot{\varphi}^{2}}\right), \eqno(93)
$$
$$
\beta^{2} = 1, \;\; w = n = - \dot{\varphi}, \;\; p = -V +
\frac{\dot{\varphi}^{2}}2, \eqno(94)
$$
$$
\epsilon = 3H^{2} = V + \frac{\dot{\varphi}^{2}}2, \;\; \dot{H} = -
\frac{\dot{\varphi}^{2}}2, \eqno(95)
$$
$$
\ddot{\varphi} + 3H \dot{\varphi} + \frac{dV}{d \varphi} = 0. \eqno(96)
$$

Inflation occurs when
$$
\ddot{a} = a(\dot{H} + H^{2}) = \frac a3 (V - \dot{\varphi}^{2})>0.\eqno(97)
$$

At $V > \dot{\varphi}^{2}$ the potential energy density contributes
dominantly to the Hubble parameter. For our functions (92) it may
happen at large $\varphi$, and inflationary solution can be got as
expansion over the inverse powers of $\mid \varphi \mid \gg 1$:
$$
H = \sqrt{\frac V3}(1+\frac 1{12}c^2\varphi^{-2}+O(\varphi^{-4})), \eqno(98)
$$
$$
\dot{\varphi} = - \frac{cH}{\varphi} (1 + \frac{1}{3}(e - 1) c
\varphi^{-2} + O(\varphi^{-4})), \eqno(99)
$$
$$
a = \exp \left[ - \int d \varphi \;\; \varphi(c^{-1} + \frac{1}{3} (1 - e)
\varphi^{-2} + O(\varphi^{-4})) \right], \eqno(100)
$$
$$
\alpha = \frac{c}{2 \varphi}(1 + \frac{1}{3}(e - 1)c \varphi^{-2}
+ O(\varphi^{-4})). \eqno(101)
$$

Then the inflationary condition can be rewritten in terms of $c$-function:
$$
\left|\frac c{\varphi} \right| < \sqrt{3}, \eqno(102)
$$
or in terms of potential $V$:
$$
\left|\frac{dV}{d \varphi}\right |< \sqrt{3} V. \eqno(103)
$$

Eqs. (98,99,100,101) are obviously true in a so-called slow-roll
approximation which assumes friction-dominated equation of motion for
$\varphi$-field (the second-derivative term in eq. (96) is subdominant).
There are two conditions for this approximation following directly from
expansions (98 -- 101):
$$
\left| \frac {c}{\varphi} \right| \le 2, \;\;\;\; 
\left|\frac{(e - 1) c}{\varphi^{2}} \right| \le
2. \eqno(104)
$$

The first condition nearly coincides with eq. (102). From the second
condition we have
$$
\left| \frac{d^{2} V}{d \varphi^{2}} \right| \;\; \le 2V. \eqno(105)
$$

So, the inflationary solution (98, 99, 100, 101) requires certain
analytical properties from potential $V(\varphi)$ (see eqs. (103,105)).
We can put it in other terms: eqs. (98, 99, 100, 101) are valued for
large $\varphi$,
$$
|\varphi |\ge\varphi_I={\rm max}\left(\frac{|c|}2,\;\; 
\sqrt{\frac{|(e -1)c|}2}\right). \eqno((106)
$$

One can show that, within class of functions (92) satisfying the
inequality (106) for 
$$
\mid \varphi \mid \ge \varphi_{1} = {\rm const} \eqno(107)
$$
where $\varphi_{1}$ is a positive root of equation $\varphi_{1} =
\varphi_{I} (\varphi_{1})$ \footnote {For estimates, $\varphi_{1} \sim 1$.
If there are few roots in eq. $\varphi = \varphi_{I}(\varphi)$ for
$\varphi \ge 1$, then the solution (98, 99, 100, 101) can be broken for
some large $\varphi$. We do not analyse here such possibility.}, eqs.
(98, 99, 100, 101) describe a trap (for growing time) separatrix
towards which all the other dynamical trajectories \footnote {In the phase
space $(\varphi, \dot{\varphi})$, initial conditions for the classical
trajectories of eqs. (95, 96) are set up on the quantum boundary
$\epsilon = 1$ which represents a kind of ellipse (circle in case of
$V_{1}$) around the central point $\varphi = \dot{\varphi} = 0$. The
radii of this ellipse along $\varphi$ and $\dot{\varphi}$ axes are
$V^{-1}$ [1] and $\sqrt{2}$, respectively.} with initial $\mid \varphi
\mid > \varphi_{1}$ approach rapidly during their dynamical evolution
to the stable point $\varphi = \dot{\varphi} = 0$. On the inflationary
separatrix (98, 99, 100, 101) $\varphi, H$ and $\alpha$ vary slowly
while the scale factor increases nearly exponentially in time, $a \sim
\exp(Ht)$.

Eqs. (98 -- 101) break when the field reaches the point $\mid
\varphi \mid \sim \varphi_{1} \sim 1$ and the further evolution
proceeds with damping oscillations around $\varphi = \dot{\varphi} =
0$. If $d^{2}V/d \varphi^{2} > 0$ at $\varphi = 0$, then
$$
H = \frac{2}{3} t^{-2}, \;\;\;\; a \sim t^{2/3}, \;\;\;\; \varphi =
2\sqrt{\frac 23}(mt)^{-1} \sin (m(t+t_{o})) \le 1, \eqno(108)
$$
where $V \simeq m^{2} \varphi^{2}/2 \;\;$ for $ \;\; \mid\varphi\mid \le
1$, here $m$ and $t_{o}$ are constants. At this stage the Universe expands
like a pressureless medium since the average cosmological pressure is
exponentially small. The medium --- coherent oscillations of spatially
homogeneous field --- is unstable in this situation and will decay in 
particles. The result is reheating and the HFU expansion beginning.

This reheating process although producing some inhomogeneities on the
horizon scale $\sim k_{1}$, cannot damage the large scale
perturbations created already during the inflationary epoch $(k < k_{1})$.

To find the postinflationary PCP spectrum we must solve Eqs. (61) with
parameters (98, 99, 100, 101) for $\varphi > \varphi_{1}$:
$$
\nu^{''}_{k} + (k^{2} - U) \nu_{k} = 0, \eqno(109)
$$
$$
U = \frac{(\alpha a)^{''}}{\alpha a} = 2(aH)^{2} \left(1 + \frac
{\dot{H}}{2H^{2}} + \frac {(a^{3} \dot{\alpha} \dot{)}}{2a^{3} \alpha
H^{2}}\right) =
$$
$$
= 2(aH)^{2} \left(1 - \frac{1}{4} c_{2} \varphi^{-2} +
O(\varphi^{-4}) \right) =
$$
$$
= \frac{2}{\eta^{2}} \left(1 + \frac{3}{4}
c_{3} \varphi^{-2} + O(\varphi^{-4}) \right), \eqno(110)
$$
$$
\eta = \int \frac{dt}{a} = - \frac{1}{aH} \left(1 + aH \int \frac
{\dot{H} dt}{aH^{2}} \right) = - \frac{1}{aH} \left(1 + \frac{1}{2}
c^{2} \varphi^{-2} + O(\varphi^{-4}) \right), \eqno(111)
$$
where $c_{2}/c = c + 6(1 - e), \;\;$ and $\;\; c_{3}/c = c + 2(e - 1)$. The
conformal time $\eta < 0$, and initial conditions are
$$
\nu_k=\frac 1{\sqrt{2k}} \exp(-ik\eta),\;\;{\rm for} \;\; 
k \mid\eta\mid \;\gg 1. \eqno(112)
$$
Eqs. (109,110,111,112) can be solved explicitely by matching two
following solutions in the overlapping region
$$
c_{4} \varphi^{-4} < k \mid\eta\mid < 1, \eqno(113)
$$
where $c_{4}/c^{2} = (c + e - 2)(1 - e) - ef$. The first solution
assuming the left inequality (113), allows for the $U$-potential
approximation by $U = {\rm const}/ \eta^{2}$ near $k \mid \eta \mid \sim 1$
(cf. eq. (110)). Since $U \ll k^{2}$ for $k \mid \eta \mid \gg 1$, we
have for $k \mid\eta\mid \; > c_{4} \varphi^{-4}$:
$$
\nu_{k} = \frac{1}{2} \sqrt{\pi \mid\eta\mid} H^{(1)}_{\nu} (k \mid
\eta\mid) = \frac 1{\sqrt{2k}} (1 - \frac {i}{k \eta}) \left(e^{-ik \eta} +
O(\varphi^{-2}) \right), \eqno(114)
$$
where $H^{(1)}_{\nu}(x)$ is the Hankel function, $\;\; \nu = (3 + c_{3}
\varphi^{-2})/2.$ 

Under the right inequality (113), eq.(85) describes the general
solution for any $U(\eta)$. Since the integral in eq. (85) converges
sharply in time, we have for $k \mid\eta\mid < 1$:
$$
\nu_{k} = i \pi \sqrt{2} k^{-3/2} \alpha a q_{k} = - \frac{i \pi}
{\sqrt{2} \eta} k^{-3/2} \left(\frac{c}{\varphi H}\right) q_{k}, \eqno(115)
$$
where constants $q_{k}$ are the PCP spectrum (see eq. (83)).

Fitting eqs. (114,115) at region (113) gives the following spectrum of
the perturbations parametrically created outide the horizon:
$$
q_{k} = \frac{1}{\pi} \left(\frac {\varphi H}{c}\right)_{k} =
\frac {1}{\pi} \left(\frac {H^{2}}{\mid \dot{\varphi} \mid}\right)_{k} = 
\frac{1}{\pi \sqrt{3}} \left(\frac {V^{3/2}}{dV/d \varphi}\right)_{k},
\eqno(116)
$$
where $\varphi = \varphi_{k}$ at the horizon crossing $(k \eta = -1)$
can be expressed directly in terms of the wave number
$$
k = aH = a(\varphi) H(\varphi). \eqno(117)
$$
The resulting spectrum (116,117) belongs obviously to the growing
perturbation mode since only for this mode $q$-field is constant in
time outside the horizon. We have already emphasized that this
property of the growing mode is independent of any expansion law or
equation of matter state. In particular, spectrum (116,117) holds in
large scales for any microphysics processes after inflation like
phase transitions or reheating.

Sometimes, people prefer to deal with the power spectrum of density
perturbations. Below, relation between $q$-scalar and the coupled
density perturbation field is obtained in a general form. 

Indeed, eq. (44) allows for the growing mode general solution outside horizon:
$$
q = q(\vec{x}) + Q(t) \Delta q(\vec{x}),
$$
$$
\dot{Q}(t) = - \left(\frac{\beta}{a}\right)^{2} \frac{H}{\dot{H}} 
(1 - \frac{H}{a} \int a dt). \eqno(118)
$$

So, the comoving density perturbations are (see eq. (52)):
$$
\delta = \frac{\delta \tilde{\epsilon}}{e}=-\frac{2 \alpha^{2}\dot{q}}{3H} 
= \frac{1}{3}\frac{1}{(aH)^{-2}}(1-\frac{H}{a} \int a dt) \Delta q(\vec{x}). \eqno(119)
$$

Now, it is not a problem to get the desired relation between power
spectra:
$$
\delta_k=\frac {q_k}3 \left(\frac k{aH}\right)^2 \left(1 - 
\frac{H}{a}\int a dt\right) = q_{k} O\left(\left(\frac{l_{H}}{l_{k}}
\right)^{2}\right) \ll q_{k}, \eqno(120)
$$

We return now to eq. (117) which gives us the connection between $k$
and $\varphi$ on the inflationary separatrix (98,99,100,101). This
equation can be solved explicitly for a class of so-called smooth
potentials $V(\varphi)$.

Let us introduce smooth potential functions $V(\varphi)$ for which
$c(\varphi)$ varies even slower than $\varphi$ \footnote {Physically,
the characteristic scales of smooth potentials are not much shorter
than $\varphi$.}:
$$
\mid e(\varphi) \mid \ll 1. \eqno(121)
$$

For such potential eq. (100) yields
$$
a \simeq \varphi^{-1/3} \exp \left(-\frac{\varphi^{2}}{2c}\right),
\;\;\;\; \varphi \ge \varphi_{1}, \eqno(122)
$$
which shows that $\varphi$ varies logarithmically in the conformal time
(cf. eq. (111)). If $c_{1} = c (\varphi_{1}) \ge 2$, then $\varphi_{1}
= c_{1}/2 \ge 1$ (see eqs. (106,107)) and
$$
\varphi^{2} = \frac c2 \left(c_{1} + 4\ln \left(\frac{\eta H
\varphi^{1/3}_{1}}{\eta_{1} H_{1} \varphi^{1/3}} \right)\right), \eqno(123)
$$
for $\varphi \ge \varphi_{1} \;\; (\mid \eta \mid \ge \mid \eta_{1}
\mid)$. The substitution to eq. (117) gives for $k \le k_{1}$: 
$$
\varphi^2_k=\frac c2 \left(c_1+4\ln\left(\frac{k_1H\varphi^{1/3}_1}{kH_1
\varphi^{1/3}}\right)\right)\simeq 2c\ln(k_1/k), \eqno(124)
$$
(the second equality implies $k \ll k_{1}$).

So, smooth potentials generate the Harrison-Zeldovich types of spectra
(see eq. (116)) growing only logarithmically to large scales. 

Typical example of smooth potential is a power-law potential
$$
V_{n} = \frac{1}{2n} a^{2}_{n} \varphi^{2n}
$$
which generates the following spectrum:
$$
q_{k} = (4 \pi n^{3/2})^{-1} a_{n} \varphi^{n+1} =
$$
$$
= \frac{2}{\pi} (4n)^{(n-2)/2} a_{n} \left [n/2 + \ln
\left(\frac{k_{1}}{k} \left(\ln\frac{k_1}{k}\right)^{(3n-1)/6} \right)
\right]^{(n+1)/2}, \eqno(125)
$$
where $a_{n}$ and $c = c_{1} = 2n \ge 2$ are constants.

Important cases are the massive field $(n = 1, \; a_{1} = m)$:
$$
q_k=\frac m{\pi}\ln\left(\frac{k_1}{k}\left(\ln\frac{k_1}{k}\right)^{1/3} 
\right), \eqno(126)
$$
and the $\lambda$-field $(n = 2, \; a_{2} = \sqrt{\lambda})$:
$$
q_k=\frac{2\sqrt{\lambda}}{\pi}\left[\ln \left(\frac{k_1}{k}\left(\ln
\frac{k_1}{k}\right)^{5/6}\right)\right]^{3/2}. \eqno(127)
$$
For non-smooth $V(\varphi)$ spectrum $q_k$ can have, in principle,
any form depending on given potential and the first derivative shapes.
Moreover, it is possible to inverse the problem and to find potential
$V(\varphi)$ for any \footnote{Some spectra may violate the slow-roll
conditions which makes the inverse problem self-inconsistent.}
given postinflationary PCP spectrum (Hodges $\&$ Blumenthal 1989). 
True, some potentials appear to be rather exotic ones, but the result 
is very important: PCP spectra are very sensitive to the potential forms 
\footnote {Physically, non-Harrison-Zeldovich spectra appear from 
potentials which have characteristic scales less then $\varphi$.}.

We shall return to the latter problem in the fifth Chapter. But now,
let us stress two more points in the conclusion.

Postinflationary perturbations (116,117) are Gaussian with random
spatial phases since it is the seed point-zero vacuum fluctuations (of
the $q$-field) from which they were parametrically created, that are
Gaussian by definition. Here, we have no problem with initial
conditions for the $q$-scalar because they are determined by
microphysics inside the horizon. 

Another interesting point is that most spectra grow with scale growing.
It means that there exists some critical field (and, thus, the critical
scale) for which the corresponding amplitude $q_{k} \sim 1$:
$$
\left(\frac{\varphi H}{c}\right)_{cr} \sim 1, \;\;\;\; k_{cr} 
\sim (aH)_{cr} \ll k_{1}.
\eqno(128)
$$

Say, for potentials (125,126,127) we have

\begin{tabular}{lll}
$n\ge 1:$ & $(\varphi H/c)_{cr}\sim 1,$ & $k_1/k_{cr}\sim\exp(
            a_n^{-2/(n+1)}),$ \\
$n=1:$    & $\varphi_{cr}\simeq 2\sqrt{\pi}m^{-1/2},$ &
            $k_1/k_{cr}\simeq(m/\pi)^{1/3}\exp(\pi/m),$\\
$n=2:$    & $\varphi_{cr}\simeq 3\lambda^{-1/6},$ &
            $k_1/k_{cr}\sim\exp(\lambda^{-1/3}).$\\
\end{tabular}

We shall see in the next Chapter that the Universe on large scales, $k
\le k_{cr.}$, is globally non-linear and it is stochastic (dominated by
quantum fluctuations) for $\varphi > \varphi_{cr.}$.

\section{Inflation}
There is no secret that Inflation is a corner stone of the VEU theories. 
This is not a surprise since up to now we have no alternative to the 
Inflationary Paradigm.

We are not going to discuss the Paradigm here. There is a lot of reviews 
and courses devoted to the subject (e.g. see Colb $\&$ Turner 1989, and 
references therein). Instead, we would like to dwell on the chaotic 
inflation which, in our view, is the first theory of the kind that can be 
called the cosmologically standard theory. At least, in a sense as this 
status has the standard Friedmann cosmology or the parametric amplification 
theory. All of them, based on simple cosmological postulates which are not
directly related to any particular particle physics, can explain and predict 
a lot of obvservational consequences (see the Introduction).
       
The goal of inflationary theory is to prepare initial conditions for
the standard FU. There are the following five items among them.
\begin{itemize}
\item[(i)] Homogeneity and isotropy along with the Euclidean geometry of the
spatial slice on scales near the contemporary horizon.       
\item[(ii)] The amplitude $\sim 10^{-4}$ of PCPs at this slice on galactic to
supercluster scales.
\item[(iii)] Reheating sufficient for the primordial entropy production,
nucleosyntesis and baryogenesis met in Friedmann cosmology. 
\item[(iv)] Small particle numbers $(\Omega \le 1)$ of the
unwanted massive relics created by the Big Bang and primordial reheating.
\item[(v)] Small density of the Friedmann vacuum which is the $\Lambda$-term
$(\Omega_{\Lambda} < 0.7)$.
\end{itemize}

First inflationary models were rather connected to the specific
physical theories and hypotheses like GUTs, phase transitions,
quantum-gravity effects, etc. However, in view of absence of the true
high energy physics and,which is more important, taking into account a
purely cosmological status of the first three items above, there was
an understanding of the necessity in constructing a cosmological
standard inflationary theory which could be independent of any current
speculations about future fundamental physics, on one side, and could
account for the first three puzzles of FU, on the other side.
Certainly, such a theory would not solve the fourth an fifth problems
which were much more related to the particle physics indeed. 

The first theory of such type was proposed by Linde (1983). A basic
assumption is that potential energy of inflaton $\varphi$-field grows
with $\varphi$ growing (see eq. (92)). The word 'chaotic' minds the
requirement for a large value of the initial $\varphi$-field $(\varphi
\gg 1)$ which could be realized somewhere in spacetime under hypothesis
of the chaotic initial conditions. However, we do not think that the
latter requirement is somehow a problem for the theory at all.

Below, we dwell on the necessary and sufficient conditions for chaotic
inflation and then discuss some implications related to the subject.

\subsection{Chaotic Inflation}
Let us dwell on Lagrangian (66) with the potential term of type (92). To
start inflation, the latter must predominate at $\varphi \gg 1$: 
$$
V(\varphi) > \mid\varphi_{,i} \varphi^{,i}\mid. \eqno(129)
$$

Let us estimate the size of the region where eq. (129) is initially met.

As we have seen in the previous Chapter the time derivative of initial
$\varphi$ is not a problem regarding the inequality (129) if the
spatial homogeneity is postulated, since the inflationary solution is a
trap separatrix for $\varphi \gg 1$. So, of principal importance is the
spatial gradient condition following from eq, (129):
$$
\mid\nabla \varphi\mid \le H. \eqno(130)
$$

Eq. (130) can be read as follows: to start inflation one has to prepare
a quasi-homogeneous distribution of $\varphi$ on scale $\sim L =
\varphi/ \mid\nabla \varphi\mid$ which is much larger than the
horizon scale:
$$
L \ge l_{SI} = \varphi H^{-1} \gg l_{H}. \eqno(131)
$$

The start inflation scale has a physical meaning of the Compton scale
of inflaton which becomes explicitly clear in case of the massive
field $(V = V_{1})$: 
$$
l_{SI} = m^{-1}.
$$

Taking it into account for estimates, the start inflation condition
(131) is not probably a great surprise in a general case as well.

Eq. (131) deals with the initial distribution of $\varphi$-field. In
principle, one can rise a question about another (additional to eq.
(131)) start inflation condidtion, namely, about initial spatial
distribution of metric (curvature) on scales less than $l_{SI}$.
However, we will not discuss this problem here by the formal reason.
From the beginning, we decided to restrict ourselves by the case
when $\varphi$-field is the only source of the metric $g_{ik}$. The
point is that the small-scale nonlinear curvature perturbations (if
any) assume another source unrelated to the $\varphi$-field since the
latter is homogeneous on scale $l_{SI}$. So, the curvature born by the
$\varphi$-field is supposed to be homogeneous on scale $l_{SI}$, as
well as the $\varphi$-field itself. 

Eq. (131) can be interpreted in another way: initial $\varphi$-field
should be large enough so that $l_{H}$ could be small. However, more
stringent constraints for the potential comes from the slow-roll
conditions (103), (105). Let us explicitly rewrite these conditions in
terms of the PCP spectrum (106).

From eq. (103) we have the potential restrictions:
$$
V^{1/2} \le 10 q_{k}. \eqno(132)
$$

Eqs (105,116) constrict the spectrum index range:
$$
\frac{d \ln q_{k}}{d \ln k} \le 2. \eqno(133)
$$

So, $q_{k}$ cannot vary from the Harrison-Zeldovich spectrum faster
than $k^{\pm 2}$ (which is quite compatible with the market of galaxy
formation theories considered today).

Eq. (132) puts the direct observational limits on the potential
amplitude $V(\varphi)$ for $\varphi = \varphi_{k}$ within the structure
scale range $k^{-1} \sim (10-10^{4}) h^{-1}$ Mpc. Eq. $q_{k}
\sim 10^{-4}$ evidences for the weak coupling of $\varphi$-field to the
potential $V(\varphi)$ in this region.

Endeed, let us demonstrate it when the $c$-function variation along the
scale range can be negligent. In this case $V = \lambda \varphi^{c}$,
and we obtain a very small value for the coupling parameter (in the
Planck units):
$$
\lambda \sim 3c^{2} 10^{-8} \varphi^{-(2+c)} \le 10^{-8}. \eqno(134)
$$

Remember that the small coupling parameter is also required for the
large size of the $FU$-bubble (to be more than the horizon today, see below).

Eq. (132) can be used for some other important constraints, e.g., on
the reheating temperature. The radiation energy after reheating cannot 
actually exceed the inflation energy near the end of inflation: 
$$
\rho_{rad} = \frac{\pi^{2}}{30} g^{\ast} T^{4}_{RH} \le
V(\varphi_{1}) \le  V(\varphi) \le 10^{2} q^{2}_{k}, \eqno(135)
$$
where $\varphi = \varphi_{k} \ge \varphi_{1} \;\; (k < k_{1}), \;
g^{\ast}$ is the total number of massless degrees of freedom of the
thermal bath particles. Eq. (135) gives the following upper limit for
the reheating temperature:
$$
T_{RH} < \left(\frac{10}{g^{\ast}}\right)^{1/4} H^{1/2} \le
\left(\frac{300}{g^{\ast}}\right)^{1/4} q^{1/2}_k. \eqno(136)
$$

Making use the microwave quadrupole anisotropy $q_{k} \sim \Delta
T/T \sim 10^{-5}$, we have for the standard model $(g^{\ast} \sim
100): \; T_{RH} \le 10^{16}$ GeV. The latter inequality can be
confirmed with help of similar estimate for the gravitational waves
produced during inflation (we do not discuss this problem here).

Next important parameter is the Friedmann slice scale $l_{F}$, i.e., a
typical scale of the part of the Universe, created by inflation, which
can be approximated by the Friedmann model. Any scale in such
quasi-homogeneous region is described by eq. (117):
$$
k=H(\varphi)a(\varphi) = H(\varphi) \exp(-N(\varphi)) \;\;[Mpc^{-1}],
\eqno(137)
$$
where $N(\varphi) = \int H dt = N_{I} + N_{F}$ is the number of
$e$-folds of the Universe expansion from the moment when the
perturbation was at the inflationary horizon and up to now
$$
N_{I} = N_{I}(\varphi) = \int \limits^{\varphi}_{\varphi_{1}}
\frac{\varphi d \varphi}{c}, \;\;\;\; N_{F} \simeq 60. \eqno(138)
$$

After substituting $\varphi = \varphi_{cr}$ from eq.(128) we have
$$
l_F=k^{-1}_F=\exp \left(\frac{\varphi^{2}_{cr}}{c}\right) \sim 
\exp\left(\lambda^{-\frac 1{1+c/2}}\right) \sim \exp(10^{4}) 
\gg 10^{28} \; [cm]. \eqno(139)
$$

The non-linear global Cauchy-Hypersurface which develops in the result
of the chaotic inflation dynamics, is not built up yet.
Nevertheless, we see no principal difficulties to solve this
problem.The point is that the global spatial Cauchy-Hypersurface cannot
exist everywhere in spacetime: it breaks in the spacetime regions where
the $\varphi$-field reaches Planckian densities $(\varphi_{Pl} \sim
\lambda^{-1/c})$, so that the semiclassical approach becomes
self-inconsistent. We show in the next Section that the latter regions
occupy the most part of the physical volume of the Universe produced by
the chaotic inflation.

\subsection{Stochastic Theory of $q$-Field}
Let us return again to small scales $l_{k} < l_{F}$ where $q$ can be
treated as a linear quantum operator against the Friedmann background.
Equation of motion of $q$-field is
$$
\ddot{q} + 3n H \dot{q} - \frac 1{a^{2}} \Delta q = 0. \eqno(140)
$$
where $n = 1 + 2 \dot{\alpha}/{3 \alpha H}, \; \alpha$ and $H$
are the classical background functions, see eqs. (93, 94, 95, 96, 98,
99, 100, 101). As we know from the previous Chapters, the large scale 
perturbations are classical for $l_{k} \gg l_{H}$ while the quantum 
perturbations affect only small scales, $l_{k} \le l_{H}$. Let us separate
these two parts of $q$-field at the inflation period assuming that $q$ is
generated by quantum perturbation:
$$
q = \Phi + F, \eqno(141)
$$
where $\Phi$ is the classical large scale part of the $q$-field operator.

To make this separation explicit let us introduce a notion of the
miniuniverse (MU) as a part of the actual space-time of the size
proportional to the horizon:
$$
l_{MU} = \zeta H^{-1} \ge l_{H}, \eqno(142)
$$
where $\zeta = {\rm const} \ge 1$. Evidently, MUs do not expand with the
comoving volume.

Now, we can define the classical $\Phi$-field as the mean value of $q$
in MU:
$$
\Phi = \Phi(t, \vec{x}) = \int K_{\sigma} (\vec{x} - \vec{x}^{'}) q
(t, \vec{x}^{'}) d^{3}  \vec{x}^{'}, \eqno(143)
$$
where $K_{\sigma} (\vec{r}) = (2 \pi)^{-3/2} \sigma^{-3} \exp (- r^{2}/2
\sigma^{2})$ is the Gaussian MU-window, and
$$
\sigma = \frac{l_{MU}}{a} = \frac{\zeta}{aH} = -n_{1} \zeta \eta
$$
is the MU-dimension in the comoving $\vec{x}$-space, $n_1=(-a\eta H)^{-1}$.

Similarly to $\Phi$ we can define the classical part of the $q$-field 
momentum:
$$
V = V(t,\vec{x}) = \int K_{\sigma}(\vec{x} - \vec{x}^{'}) \dot{q} (t, 
\vec{x}^{'}) d^{3} \vec{x}. \eqno(144)
$$

The evolution of the coarse grained fields $\Phi$ and $V$ is governed
by the quantum perturbations presented by the $F$-operator: in the
inflationary process new and new perturbations created inside the
horizon inflate, one followed by another, outside the horizon and
start contributing to the classical fields $\Phi$ and $V$ when their
scales become about (and then larger) than $l_{MU}$. So, $F$ plays a
role of the stochastic generator for $\Phi$, the latter moving like a
Brownian particle in the gas. More of this, the dynamical equations are 
similar as well. 

Let us present eqs. (143,144) as the Fourier integrals (see eq. (61)):
$$
\Phi = \int d^{3} \vec{k} \;\;\Theta (a_{\vec{k}} q_{\vec{k}} + 
a^{\dag}_{\vec{k}} q^{\ast}_{\vec{k}}),
$$
$$
V = \int d^{3} \vec{k} \;\;\Theta (a_{\vec{k}} \dot{q}_{\vec{k}} +
a^{\dag}_{\vec{k}} \dot{q}^{\ast}_{\vec{k}}), \eqno(145)
$$
where
$$
\Theta = \Theta (k \sigma) = e^{-\frac{1}{2} k^{2} \sigma^{2}} = \int
K_{\sigma} (\vec{r}) e^{i\vec{k} \vec{r}} d^{3} \vec{r}.
$$
Then the original eq. (140) can be rewritten in terms of the classical fields:
$$
\dot{\Phi} - V = f,
$$
$$
\dot{V} + 3n HV = \frac{Hg}{\zeta^{2}}, \eqno(146)
$$
$f$ and $g$ can be called the noise functions (or generators) driven by
the quantum fluctuations:
$$
f = mH \sigma^{2} \int d^{3}\vec{k} \;\;k^{2} \Theta(a_{\vec{k}} q_{\vec{k}} +
a^{\dag}_{\vec{k}} q^{\ast}_{\vec{k}}),
$$
$$
g = H \sigma^{2} \int d^{3} \vec{k} \;\;k^{2} \Theta (a_{\vec{k}}
p_{\vec{k}} + a^{\dag}_{\vec{k}} p^{\ast}_{\vec{k}}), \eqno(147)
$$
where
$$
p_{\vec{k}} = - q_{\vec{k}} + \frac{m \zeta^{2}}{H} \dot{q}_{\vec{k}}, 
\;\;\;\;m = - \frac{\dot{\sigma}}{\sigma H}.
$$

Before we calculate correlators of the noise functions let us introduce
the normal modes of the classical fields,
$$
\tilde \Phi = \Phi + \frac{m_{1}}{3H} V, \;\;\;\; \tilde V = V,
$$
and separate the Eqs. (146):
$$
\dot{\tilde \Phi} = \tilde f. \eqno(148)
$$
(Eq. (146) for $V$-function did not change). Here $\tilde f = f +
m_{1}g/ \zeta^{2}.$ The background functions $n_{(1)}$ and $m_{(1)}$ are
easily derived on the inflationary separatrix $(\varphi \gg 1)$: 
$$
n = 1 + c(1 - e) \varphi^{-2} + O(\varphi^{-4}),
$$
$$
n_{1} = 1 - \frac{1}{2} c^{2} \varphi^{-2} + O(\varphi^{-4}),
$$
$$
m = 1 - \frac{1}{2} c^{2} \varphi^{-2} + O(\varphi^{-4}),
$$
$$
m_{1} = 1 - \frac{1}{6} c^{2} \varphi^{-2} + O(\varphi^{-4}). \eqno(149)
$$

Eq. (148) coincides with the Langeven equation describing the drift of
a Brownian particle if the particle coordinate is understood instead
of $\Phi$.

When comparing windows $\Theta$ and $k^{2} \Theta$ for the classical
and noise functions, we can see that in the latter case the main
contribution comes from scales $l \sim l_{MU}$. It is clear: the field
averaged over the mini-universe, can change its value not before the
new perturbation reaches the $MU$-scale which happens in a characteristic
time (step-time) $\Delta t \sim l_{MU}$. Since the perturbations phases
appearing on the $MU$-scale are random, the process of the classical
field change is stochastic. To calculate the characteristic values of
this process we must know the correlators of the noise functions on the
Friedmann hypersurface $t = {\rm const}$.

If the $q$-field is in the vacuum state then
$$
\langle a^{\dag}_{\vec{k}} a_{\vec{k}^{'}} \rangle = 0, \;\;\;\;
\langle a_{\vec{k}} a^{\dag}_{\vec{k}^{'}} \rangle = \delta (\vec{k} -
\vec{k}^{'}),
$$
and the process is the Gaussian one, so that, the second correlators
are quite sufficient to know about. In this case the perturbation
amplitudes for $\varphi > 1$ and $\zeta > 1$ are as follows (see eqs.
(61,115,118)): 
$$
q_{\vec{k}} = i \sqrt{2} \left(\frac{H \varphi}{c}\right)_{k} 
\exp (i\vec{k} \vec{x}), \;\;\;\;
\frac{\zeta^{2}}{H} \dot{q}_{\vec{k}} = k^{2} \sigma^{2} q_{\vec{k}}.
$$

Finally, after the straightforward calculation we have in the main
approximation over $\varphi$ and $\zeta$:
$$
\langle\tilde f\tilde f^{'}\rangle=
\langle ff^{'}\rangle = 
\frac 12(\langle fg^{'}\rangle+\langle gf^{'}\rangle) = 
\langle gg^{'}\rangle=2D\delta(t-t^{'}), \eqno(150)
$$
where $D = D(t) = H^3(\varphi/2 \pi c)^2$ is the diffusion coefficient. 
The $\delta$-function in eq. (150) is used instead of each of the following 
expressions,
$$
2H\left(\frac{\sigma}{\sigma^{'}} + \frac{\sigma^{'}}{\sigma}\right)^{-2} 
\;\;\; {\rm and} \;\;\; 12H \left(\frac{\sigma}{\sigma^{'}}+\frac{
\sigma^{'}}{\sigma}\right)^{-4}, 
$$
because the halfwidths of the latter bell-functions are about the
cosmological horizon $(\Delta t \sim H^{-1})$ which is less (by the
$\zeta$-factor) than the $MU$-scale.

Now, we can introduce the probability distribution $P = P(t, \tilde
\Phi)$ to find field $\tilde \Phi$ at time $t$. By definition,
$$
\int P d \tilde \Phi = 1.
$$

Following the standard methods, we can derive the Fokker-Planck
equation for this function:
$$
\frac{\partial P}{\partial t} = D \frac{\partial^{2} P}{\partial
\tilde \Phi^{2}}. \eqno(151)
$$
Obviously, the field dispersion grows in time in this stochastic process.

Let us take some arbitrary $MU$ at time $t_{o}$ with the classical
field $\tilde \Phi_{o}$, which we call the mother. During inflation
the physical spatial volume which belonged to the mother $MU$ at $t
\sim t_{o}$, expands to larger and larger scales. For $t > t_{o}$,
this volume can be covered by other $MUs$ (daughters) \footnote {At
each step $\Delta t \sim \zeta H^{-1}$ the mother volume expands by
factor $N \sim \exp (3 \zeta)$, so there are about $N$ daughters of the
first generation, $N^{2}$ of the second, and so on so forth, inside the
volume.}. The $\tilde \Phi$-field varies from one daughter to another,
and the r.m.s. deviation (from $\tilde \Phi_{o}$) $\tilde \sigma =
\tilde \sigma (t)$ can be calculated making average either by the
quantum $q$-state in one $MU$ or over the daughter $MUs$ assembly:
$$
\tilde \sigma^{2} = \langle (\tilde \Phi - \tilde \Phi_{o})^{2}
\rangle = \int (\tilde \Phi - \tilde \Phi_{o})^{2} P d \tilde \Phi =
$$
$$
= 2 \int D dt = - \frac{1}{2 \pi^{2}} \int \frac{V^{4}d
\varphi}{(dV/d \varphi)^{3}} \sim t. \eqno(152)
$$

It can be confirmed also by the exact solution of eq. (151)
$$
P(t, \tilde \Phi) = \frac{1}{\sqrt{2 \pi} \tilde \sigma} \exp \left(-
\frac{(\tilde \Phi - \tilde \Phi_{o})^{2}}{2 \tilde \sigma^{2}} \right).
\eqno(153)
$$

At the beginning
$$
\tilde \sigma (t_{o}) = 0,\;\;\;\; P(t_{o},\tilde \Phi) = \delta(\tilde
\Phi - \tilde \Phi_{o}),
$$
while, during time, the distribution (153) broadened around $\tilde
\Phi_{o}$ with equal probability for both signs of the deviation
$(\tilde \Phi - \tilde \Phi_{o}).$ The typical one-step-change of the
$\tilde \Phi$-field is 
$$
\Delta(\tilde \Phi) = \tilde \sigma(\Delta t) = \sqrt{\frac{\zeta D}H}
\sim \sqrt{\zeta} H. \eqno(154)
$$

Before we discuss some implications of this stochastic process, let us
consider the necessary conditions for the diffusion approach.

Eqs. (148,150) are generally true if the $D$-function varies slower
than the characteristic step-time $\Delta t \sim \zeta H^{-1}$. This
requirement is commonly satisfied on the  inflationary separatrix.   

The Fokker-Planck approach (151) is less reliable here. Indeed, the
notion of the $P$-function assumes that its characteristic change-time
should be no larger than the step-time. However, regarding eq. (153),
it can be marginally so if $\zeta$ is not too high. Below, we will
assume that $\zeta \sim 1$.

\subsection{Non-Linear Inflation}
Mini-universes of size $H^{-1} (\zeta \sim 1)$ introduced in Section
(4.2), are just the causally connected regions of the inflating space.
If any two points with constant comoving $\vec{x}$-coordinates (i.e.,
expanding with the Universe) belong initially to the same $MU$ then
they will manage to exchange the light signals at least once. But if
they belong to two different $MUs$ then the light signal sent from one
point will never get the other one. An important consequence is that
each $MU$ expands in time independently of any other.

It means that any $MU$ can be chosen as mother regarding the next
generations of the daughter $MUs$ it produces. In its turn, any
daughter taken at a moment $t \sim t_{o}$, although created by some
mother at $t < t_{o}$, is the mother itself for $t > t_{o}$. This
picture has neither beginning no end. Actually, this boiler of $MUs$ is
eternally self-reproducing inflationary Universe.

Let us consider some physical volume expanding with the comoving space.
New and new $MUs$ are created inside the volume during the evolution.
We can connect by the time-like tracks causally related $MUs$
(mother-daughter, mother-daughter, etc.). We saw in the previous
Section that there is an equal probability for both signs of the field
deviation $\Delta \tilde \Phi$ to be found along any track from the
past to future. Since any $MU$ develops independently of the previous
history and its neighbours, we can forget about the seed mother field
$\varphi^{(o)}$ and try to find the current classical quasi-Friedmannian field
$\varphi_{MU}$ driven the given $MU$ on the track and applicable only
to this $MU$. Certainly, $\varphi_{MU}$ is the local $\varphi$-field
renormalized each time by the classical part of $q$-field (see eq. (141)).

Technically, we can use Newtonian gauge to find $\varphi_{MU}$, since
this particular frame most closely imitates the local Friedmannian
expansion (see eqs. (53)):
$$
ds^{2}_{MU} = \left(1 - \frac Ha P_{MU}\right) dt^{2} - a^{2} 
\left(1 + \frac Ha P_{MU}\right)dx^{2}, \eqno(155)
$$
where
$$
P_{MU} = \int a \gamma \Phi dt.
$$

The local time is
$$
t_{MU} = t - \int \frac H{2a}P_{MU} dt.
$$
and the local expansion factor is
$$
a_{MU} = a\left(1 + \frac H{2a} P_{MU}\right).
$$

In both equations we disregarded the dependence of the $P_{MU}$
function on the $\vec{x}$-coordinates within $MU$. In doing so, we can
easily recover the $\varphi_{MU}$-field from background eqs. (95). In
the main approximation over $\varphi \gg 1$, we have:
$$
\frac{\varphi_{MU}}{ \varphi^{(o)}} = 1 + \frac 1c\left(\gamma \Phi - 
\frac HaP_{MU}\right). \eqno(156)
$$

Below, we give only qualitative ideas about some results of this 
investigation.

The most interesting question arisen is as to which densities of the
$\varphi$-field most tracks lead during stochastic inflation? The
answer is a production of two factors: the probability to find a
certain $\varphi$-field on one track, and the number of tracks carring
given $\varphi$-field. As far as the first factor is concerned, the
classical monotonic decrease of $\varphi$ when sliding down the
potential $V(\varphi)$, is the smaller the larger the field is: $\Delta
\varphi_{cl} \sim - c/\varphi$ for one step-time $\Delta t \sim
H^{-1}$. On the other hand, the field stochastic change due to the
quantum perturbations proceeds in both directions of $\varphi$ with
amplitude $\mid \Delta \varphi_{st} \mid \sim H$ growing to higher
field values. So, for large enough $\varphi (\varphi > (c^{2}/
\lambda)^{1/(2+c)})$ the quantum stochastic process predominates, so
that, at each step half of the created daughters have higher $\varphi$
than their mother. But the total number of tracks created per unit
time grows to higher fields as well, $dN/dt \simeq 20H$. So, the
majority of tracks leads to high densities, thus, the largest part of
global physical volume is occupied by the Planckian field density,
i.e., by the space-time foam.   

We can only guess what is happenning there, in these most typical high
density states of the global Universe, --- the notions of the space-time
and inflaton break, mutable transitions to different physics,
signatures, dimensions and other conceivable and inconceivable
worlds may occur. In fact, we can only say that the inflation states
with densities less than the Planckian one, are non-typical and very
unprobable ones in this really chaotic Universe dominated by the sea of
quantum fluctuations. The stochastic regime considered above is just a
part, the semiclassical part of this sea, where the space-time is
already classical while the inflaton is still dominated by quantum
fluctuations. Evidently, such {\it regions} decouple  occasionally from the
space-time foam and exist independently during some period of the
{\it classical time}.

A very important conclusion is that inside these semi-classical
regions there exist some very few tracks which lead occasionally
(through the random stochastic process) to lower and lower densities of
$\varphi$-field. When the latter becomes below the critical quantity
$(\varphi \le (c^{2}/ \lambda)^{1/(2+c)})$, quantum fluctuations are
not able any more to increase $\varphi$ in the created daughter- $MUs$
and the successive inflation continues with a monotonic decrease of the
inflaton. Now, the quantum fluctuations are responsible only for small
density perturbations varing slightly from one daughter to other. So,
the result will be the adiabatic Gaussian perturbations (with the
amplitude decreasing to smaller scales) against the Friedmannian
background patch surrounded by the non-linear chaotic Universe.

Let us emphasize two other points in the conclusion.

The global Cauchy-Hypersurface does not rigorously exist. It can be
constructed near the Friedmannian patch in a space-time region
restricted by the Planckian densities. In fact, such a non-linear
solution describes just a temporal island formed with a very small
probability in the chaotic space-time foam of the Universe and suitable
for life. In this connection, we can mention that we do not think it
is worth while putting seriously the question as to how general are the
start-inflation conditions for the chaotic or other inflation theories
from the point of view of general solutions of the CGR equations? In
our opinion, it is quite enough that the probability for creating a
low-density world where life can appear, is non-zero. We have no time
to  discuss this subject in more detail here.

The next important point is as follows. There are the inflaton quantum
fluctuations in the chaotic inflation theory that are the reason for
both, the non-linear global structure of the Universe as a whole and
the adiabatic density perturbations responsible for the large scale
structure formation in the Friedmannian patch of such a Universe.
Therefore, we can test the inflationary theories just investigating the
spectrum of PCPs and then reconstructing the global structure within
the theory frameworks. This is, probably, the only informative
cosmological channel to learn anything about the features beyond
optical horizon, as well as the fundamental physics parameters beyond
direct experiment. We can see that the problem of testing inflation becomes
part and parcel of the VEU theories. We are going to discuss it briefly in the
next Chapter.

\section{Testing VEU}
There are few important cosmological predictions coming from the very
early inflationary epoch. Among them are the total energy density in
the Universe and the spectrum of adiabatic PCPs.

The former quantity is to be equal to critical density up to accuracy
of the PCP amplitude on the contemporary horizon:
$$
\Omega_{tot} = 1 \pm \delta_{H},\;\;\;\;\mid K \mid\le\delta_{H}.\eqno(157)
$$ 

This amplitude $\delta_{H} = \delta_{k} (k = H_{o})$ can be easily
estimated by the quadrupole anisotropy of the microwave
background radiation:
$$
\delta_{H} \sim 10^{-4}.
$$

The real dynamical density is close to the critical one with the
accuracy $\sim 30 \%$. However, even if eq. (157) were confirmed with a
much higher accurary by future observations, it could not, unfortunately, 
tell us much information about the inflation principal parameters. 
In this respect, more informative is the PCP spectrum generated by 
last stages of the inflationary epoch before the beginning of FU.

As far as the chaotic inflation is concerned, the postinflationary PCP
spectrum is very sensitive to the potential form (see eqs. (116, 117).
On the other hand, the shape of the potential energy of the inflaton
found in a given Particle Physics, must be unambiguously  fixeds by the
fundamental Lagrangian regarding all the particle fields and interactions.

Thus, a principle test of inflation is the large scale structure of
the Universe. Its analysis allows for restoring of the
postrecombination PCP spectrum on galactic to horizon scales. This part
of the spectrum is obviously related to the postinflationary PCPs
which, in their turn, depend directly on the inflaton potential within
the field interval responsible for the scales mentioned above (see eq. (117)).

Two following questions arise from this consideration:
\begin{itemize}
\item[$\ast$] How to relate the postinflationary and postrecombination 
spectra?
\item[$\ast \ast$] Which are the basic cosmological  observations now pouring
the light on the PCP spectrum?
\end{itemize}

Below, both topics are discussed very briefly.

\subsection{Transfer Functions}
The point is that any real confrontation of theory with observations
can be done only within the framework of some cosmological model
allowing to transfer PCPs from times when they were produced and up to
the moment when they entered the non-linear evolution to form the
hierarchy of the objects observed. The principal parameters of the
model are those of the dark matter components running out the
gravitational evolution of PCPs.

Regarding the gravitational impact to PCPs there are two components of
the Dark Matter Cold (CDM) and Hot (HDM) that are important. The cold
particles may be heavy relics $(m_{C} \gg 10 eV)$ or coherent axions 
which behave like a non-relativistic medium.
The hot particles are those like massive neutrino with the equilibrium
particle density and according restmass $(m_{H} \le 10 \; eV)$. These
two components evolving very differently in the past are both
non-relativistic and maintain the critical density now:  
$$
\Omega_{C} + \Omega_{H} + \Omega_{\Lambda} = \Omega_{tot} = 1. \eqno(158)
$$
Baryons can be added in $\Omega_{C}$, the third term which is the
energy of physical vacuum cannot be totally excluded today 
$0<\Omega_{\Lambda}< 0.7$.

The rest dark matter, which we call as the $\nu$-component, does not
contribute  crucially to eq.(158) and consists of relativistic and
semirelativistic weakly interacting particles $(m_{\nu} \ll 1 eV)$.
We can characterize $\nu$-particles by there total number $N_{\nu}$
with respect to the relic photons:
$$
\nu = \frac{N_{\nu}}{N_{\nu} + N_{\gamma}} \in (0,1). \eqno(159)
$$

For the standard CDM and HDM models $\nu = 0.4$ and $\nu = 0.3$
respectively, counting three or two sorts of the massless neutrino.
Generally, $\nu$-particles include gravitions, light SUSY and other
hypothetical ...inos probably existing in the Universe. 

So, in the simplest matter dominated case $(\Omega_{tot} = 1, \; 
\Lambda = 0$, stable
particles) we have  two free parameters, $\Omega_{H}$ and $\nu$, both
ranging from zero to one, which determine the past history of PCPs
beginning from Inflation. The goal is to find the ratio of the final to
initial PCP spectra as a function of these two (or more in a general case) 
parameters. This ratio is called the transfer function,
$$
T(k) = \frac{q^{(f)}_{k}}{q^{(i)}_{k}}, \eqno(160)
$$
where postinflationary spectrum $q^{(i)}_{k}$ coincides with the function 
$q_{k}$ from eqs. (116, 117), $q^{(f)}_{k}$ is the postrecombination PCP 
spectrum responsible for LSS formation in the Universe. Evidently, $T(k)$ 
does not depend on the inflationary period, it is a functional of the 
Friedmann model dynamics from the beginning to our days.

$T(k)$ is equal to unity for very large scales $(T(0) = 1)$ and then
decreases monotonically with $k$ growing. It still remains to be about
unity up to some characteristic scale $l_{eq}$ which coincides with the
horizon at period at equality of all relativistic and all
non-relativistic component densities. The further $T(k)$-fall-shape to
shorter wavelengths is an intrinsic property of the model (we refer
this subject to the special courses and lectures).

For the standard models $l_{eq} \sim 30h^{-2}$ Mpc but, for arbitrary
$\nu, l_{eq}$ grows with $\nu$ growing like $l_{eq} \sim (1- \nu)^{-1/2}.$

As we see, the resulting spectrum $q^{(f)}_{k}$ is a sensitive
function of both fundamentals of the early Universe, the
postinflation PCPs and dark matter composition. So, the
investigation of its direct creature --- the large scale structure of the
Universe --- cannot be overestimated today. Another principle test --- a
laboratory detection of the dark matter particles --- is not discussed here.

\subsection{Observations}
Since there are few special courcses devoted to this subject we only
briefly outline the hot spots of the confrontation between observations
and theory important for us. We see tremendous importance for
the modern cosmology of two groups of experiment nowdays: $\Delta T/T$
observations both on large and small scales $(\Theta^<_> 1^{o})$ and direct
observations of the distribution and evolution of the hierarchies of LSS.

The point is that both experiments confront and complement each other.

If $\Delta T/T$ upper limits and detections which become available 
now, make us to lower down the primordial perturbation amplitudes on scales 
larger than $l > 10 h^{-1}$ Mpc, then LSS needs for its existence high
enough cosmological perturbation amplitudes on scales $l\sim 10 - 100
h^{-1}$ Mpc. For the most theories of galaxy formation the gap between
these two requirements is quite negligible. Say, within the Gaussian 
perturbation theories any reasonable asumption for large superclusters and
voids to be more or less standard phenomenon in the visible Universe, leads
inevitably to the $\Delta T/T$ prediction levels on degrees of arc capable for
current detection. It brings a very great optimism to obtain large scale
primordial perturbation spectrum directly from the observations
with a high degree of accuracy.

The current situation with $\Delta T/T$ is well known. For our case of
Gaussian PCPs we may directly relate our scalar $q$ on the last
scattering surface at recombination with the map of the temperature
anisotropies on the selestial sphere $\vec{n} = \vec{n} (\theta,
\varphi)$. Endeed, let us decompose the latter in spherical functions:
$$
\frac{\Delta T}{T} (\vec{n}) = \sum \limits_{l, m} a_{lm} Y_{lm}
(\theta, \varphi) . \eqno(161)
$$

Then, after simple calculations, the temperature correlation function
takes the following form:
$$
C(\alpha) \equiv \langle \frac{\Delta T}{T} (\vec{n}_{1}) \frac{\Delta
T}{T} (\vec{n}_{2}) \rangle = \sum \limits_{l} C_{l} P_{l}
(\cos \alpha), \eqno(162)
$$  
where $\vec{n}_{1} \vec{n}_{2} = \cos \alpha, \;\; P_{l} (\cos \alpha)$
are the Legandre polinomials, $\langle...\rangle$ is the average over 
the field state, and
$$ 
C_{l} = \frac{2l + 1}{4 \pi} a^{2}_{l} =
\frac{1}{4 \pi} \sum \limits_{m} \langle a^{2}_{lm} \rangle . \eqno(163)
$$

Generally, there are three main sources of the primordial temperature
anisotropies: due to fluctuations of the gravitational potential,
matter density and velocity perturbations. For the vanishing pressure
they may be reduced to the following expression taken at recombination: 
$$
\frac{\Delta T}{T} (\vec{n}) = \frac{1}{3} q (\vec{x}_{r}) +
\frac{1}{4} \delta_{\gamma, r} - \vec{n} \vec{v}_{r}. \eqno(164)
$$

The first term (Sachs-Wolf effect) dominates for $\alpha > 1^{o} (l >
100 \; h^{-1} \;$ Mpc) which yields the direct connection between
$C(\alpha)$ and the PCP spectrum $q_{k}$: 
$$
a^{2}_{l} = \frac{2}{\pi} \int \frac{dk}{k} q^{2}_{k} J^{2}_{l} (k),
\;\;\; C(\alpha) = \frac{1}{2 \pi^{2}} \int \frac{dk}{k} q_{k}^{2}
J_{o} (k \beta), \eqno(165)
$$
where $J_{l}$ are the Bessel functions, $\beta = 2 \sin \frac{\alpha}{2}$.

The most important is the COBE detection for $\alpha \sim
10^{o}$ evidencing the consistency with the HZ-spectrum on very large
scales, $l \sim 1000 \; h^{-1} \;$ Mpc: 
$$
\Delta^{2}_{k} \sim k^{4} q^{2}_{k} \sim k^{\alpha}, \;\;\; \alpha =
4 \pm 0.2 , \eqno(166)
$$
where $\delta^{2} = (1 + z)^{-2} \int \frac{dk}{k} \Delta^{2}_{k}$ is
the mean square perturbation of density.

As for the LSS data, we have many independent indications today 
for existence of the large scale structures up to a typical scale  
$l_{LS}\simeq 100\,h^{-1} Mpc$. The most important data come from 
the Great Attractor $(z\leq 0.03)$, pencil beam galactic surveys 
$(z\leq 0.3)$, existence of large groups of quasars ($z\in (0.5,2)$),
and spatial distributions of clusters of galaxies $(z\leq 0.1)$. 
These LSS experiments indicate the following estimate for the spectrum
of Gaussian density perturbations within the scales $l\in(10,100)\,h^{-1}
\,Mpc$:
$$
\Delta_k^2\sim k^\gamma,\;\;\;\;\gamma=0.9\pm 0.2. \eqno(167)
$$
The consistency with eq.(166) gives us the obvious evidence for the 
presence of a real feature
in the power spectrum at the supercluster scale $\sim 100\,h^{-1}\,Mpc$;
it may a change in the spectrum slope from HZ (at $l>100\,h^{-1}\,Mpc$)
to the flat one ($l>100\,h\,Mpc$). This "signature of the God" requires
its explanation in physics of the very early Universe.

We do not discuss here modern status of the cosmological model and
observational tests refering to special reviews (e.g. astro-ph/9803212).

\section{Conclusion}
As we have seen, still in the absence of high energy physics, we may
successfully develop the theory of VEU on purely cosmological grounds
and come to important conclusions about the spectrum of PCPs capable
of current testing by observations. On the other hand, we are very
close to recover the postrecombination PCP spectrum directly from the
observations, both $\Delta T/T$ and LSS (especially its evolutionary
aspects), and thus to reconstruct the true cosmological model and make
the exciting link to VEU physics. Both confronting branches -- theory 
and observations -- develop fruitfully and make us hope to find the 
principal answers on the evolutionary model of the Universe during our days.

This work was partly supported by INTAS 97-1192.

\section*{Appendix A}
Here we obtain the Lagrangian $L^{(2)}$ (see eqs. (26)).

Asuming that eqs. (25) are exact ones, all the auxiliary quantities are
expanded to the second order in $\phi$ and $h^{ik}$:
$$
\delta w = \frac{1}{2w^{(o)}} (\delta (\varphi_{,i} \varphi_{,k}
g^{ik}) - (\delta w)^{2}) = w^{(o)} (\chi - \frac{1}{2} \chi^{2} +
\frac{1}{2} v_{i} v^{i} - u_{i} v_{k} h^{ik}),
$$
$$
\delta L = n^{(o)} (\delta w - \nu^{(o)} \phi + \Gamma \frac{\delta
w}{w} \phi + \frac{1}{2w} ( (\frac{\delta w}{\beta})^{2} - m^{2}\phi^{2}) =
$$
$$
= n^{(o)} w^{(o)} (\chi - \nu^{(o)} v + \Gamma \chi v +
\frac{\chi^{2}}{2} (\beta^{-2} - 1) + \frac{1}{2} v_{i} v^{i} - u_{i}
v_{k} h^{ik} - \frac{1}{2} m^{2} v^{2})
$$
$$
\delta g_{ik} = h_{ik} + h_{il} h^{l}_{k}, \;\;\;\; \ln
\left(\frac{g}{g^{(o)}}\right) = h + \frac{1}{2} h^{k}_{l} h^{l}_{k},
$$
$$
\delta \sqrt{-g} = \frac{1}{2} \sqrt{-g^{(o)}}\;\; (h + \frac{1}{2}
h^{k}_{l} h^{l}_{k} + \frac{1}{4} h^{2}),
$$
$$
\delta \Gamma^{l}_{ik} = \frac{1}{2} (h^{l}_{i;k} + h^{l}_{k;i} -
h^{;l}_{ik} + h^{m}_{i} h^{l}_{m;k} + h^{m}_{k} h^{l}_{m;i} + h^{l}_{m}
h^{;m}_{ik} - (h_{im} h^{m}_{k})^{;l}),
$$
$$
\delta\Gamma^{l}_{il}=\frac{1}{2} (h + \frac{1}{2} h^{k}_{l}h^l_k)_{,i,}
$$
$$
\delta R = g^{ik} \delta R _{ik} - h^{ik} R^{(o)}_{ik},
$$
$$
\delta R_{ik} = (\delta \Gamma^{l}_{ik})_{;l} - (\delta
\Gamma^{l}_{il})_{,k} + (\delta \Gamma^{l}_{ik}) (\delta
\Gamma^{m}_{lm} - (\delta \Gamma^{m}_{il})(\delta \Gamma^{l}_{km}),
$$
where $\delta f = f - f^{(o)}, \;\; v = \phi/w^{(o)}, \;\; v_{i} =
\phi_{,i}/w^{(o)}, \;\; \Gamma^{l}_{ik}$ are the Christoffel symbols.

The substitution to eq. (20) yields:
$$
(L - \frac{1}{2} R) \sqrt{\frac {g}{g^{(o)}}} = L^{(o)} - \frac{1}{2} R^{(o)}
+\frac{1}{2} (1 + \frac{1}{2} h) (G^{(o)}_{ik} - 
$$
$$
- T^{(o)}_{ik} - \phi((n^{(o)} u^{l})_{;l} + n^{(o)} \nu^{(o)}) + S^{l}_{;l}
+ L^{(2)},
$$
$$
S^{l} = \frac{1}{2} (1 + \frac{1}{2} h) (g^{il} \delta
\Gamma^{k}_{ik} - g^{ik} \delta \Gamma^{l}_{ik}) + n^{(o)} u^{l} \phi,
$$
$$
L^{(2)} = \frac{1}{2} n w(v_{i} v^{i} + \chi^{2} (\beta^{-2} -
1) - 2v_{i} \psi^{i}_{k} u^{k} + \nu v \psi - 
$$
$$
- m^{2} v^{2} + 2 \Gamma v \chi) + \frac{1}{4} (L^{o} - \frac{1}{2}
R^{(o)}) (\psi_{ik} \psi^{ik} - \frac{1}{2} \psi^{2} ) +
$$
$$
+ \frac{1}{8} (\psi_{ik;l} \psi^{ik;l} - 2 \psi_{ik;l} \psi^{il;k} -
\frac{1}{2} \psi_{,l} \psi^{,l}).
$$

If background metric satisfies the classical equations,      
$$
G^{(o)}_{ik} - T^{(o)}_{ik} = 0, \;\;\;\; (n^{(o)} u^{l})_{;l} +
n^{(o)} \nu^{(o)} = 0,
$$
then the substitution to $L^{(2)}$ gives eq. (26). Note, that the
linear terms in $\phi$ and $h^{ik}$ (see eq. (20)) prove to be zero
since the background equations are met.

\section*{Appendix B}
Here we derive general relations between $q$-scalar and the
perturbations in arbitrary reference frame.

Let us decompose perturbations (25) over the irreducible
representations of potential type in general Friedmann model with the
metric $g_{ik}$ and 4-velocity $u^{i}$:
$$
v = X + \frac{1}{2}(C + D_{,i} u^{i}),
$$
$$
h_{ik} = Y e_{ik} + Z g_{ik} + (C u_{(i})_{;k)} + D_{;ik}, \eqno(169)
$$
where 4-tensor $e_{ik} = 2u_{i} u_{k} - g_{ik}$ has Euclidean signature
and subbrackets mean the symmetrization. Functions $X, Y, Z, C, D$ are
coefficients of the linear decomposition. $X, Y$ and $Z$ are gauge
invariant 4-scalars. $C$ and $D$ are arbitrary functions specifying the
gauge feedom of perturbations in eqs. (25) (see eqs. (28, 29)):
$$
\xi_{i} = \frac{1}{2} (C u_{i} + D_{,i}).
$$
The Einstein linear equations can be decomposed as a 4-tensor over the
irreducible representations as well:
$$
h^{;l}_{ik;l} - 2 h^{l}_{(i;k)l} + h_{;ik} + (p - \epsilon) h_{ik} +
$$
$$
+ (\epsilon + p) (\tilde{\chi} (\beta^{-2} - 1) e_{ik} + 4v_{(i}
u_{k)} - 2 \nu v g_{ik}) =
$$
$$
= E e_{ik} + F g_{ik} + (I u_{(i})_{;k} + J_{;ik} = 0,
$$
where
$$
E = Y^{;l}_{;l} - 4H Y_{,l} u^{l} + (p - \epsilon + 4(\dot{H} +
H^{2}))Y +
$$
$$
+ (\epsilon + p) (\tilde{\chi} (\beta^{-2} - 1) + 2(4H + \nu) X),
$$
$$
F = Z^{;l}_{;l} + (p - \epsilon) Z - 4H (Y_{,l} u^{l} + HY -
(\epsilon + p) X),
$$
$$
I = -2(Y_{,l} u^{l} + HY - (\epsilon + p)X),
$$
$$
J = 2Z, 
$$
and $\tilde{\chi} = w^{-1}(wX)_{,l} u^{l} - (Y + Z)/2$. Here, the
auxiliary relations
$$
u_{i;k} = H P_{ik} = \frac{1}{2} H (g_{ik} - e_{ik}), \;\;\;\; H =
\frac{1}{3} u^{l}_{,l},
$$
and some other background formulae were used. 

From $J = 0$ we have \footnote{The scalar $J$ is generally connected to the
anisotropic pressure which is zero for $\varphi$-field}:
$$
Z = 0  \eqno(170)
$$

From $I = F = 0$ the following relation between $X$ and $Y$ scalars is 
obtained:       
$$
\dot{Y} + HY = (\epsilon + p) X. \eqno(171)
$$

For the spatially flat model $(K = 0) \; q$-scalar is given by the
linear superpositions of the gauge invariant functions:
$$
q = - (Y + 2HX) \eqno(172)
$$

Obviously, it is the real 4-scalar (as well as $X$ and $Y$) independent of any
reference frame.

The inverse transformations follow from eqs. (171, 172):
$$
X=\frac 12\left(\frac Pa-\frac qH\right), \;\;\;\; Y=-\frac Ha P,\eqno(173)
$$ 
where $P = \int a \gamma q dt, \;\; \gamma = - \dot{H}/H^{2}$.

From $E = 0$ we have the key equation \footnote{Functions  $(\epsilon +
p) \tilde{\chi}$ and $(\epsilon + p)X$ should be excluded from $E$ with
help of eq. (171).}:
$$
a^{3} \gamma \beta^{-2} \dot{q} = \Delta P. \eqno(174)
$$
The rest is to prescribe the potentials $C$ and $D$ for different gauges.

Projecting eqs. (169) on the Friedmann reference system we have for an
arbitray gauge:
$$
v = X + \frac{1}{2} F, \;\;\; h_{oo} = Y + \dot{F}, 
$$
$$
h_{o \alpha} = \frac{1}{2} \psi_{, \alpha}, \;\;\;\; h^{\beta}_{\alpha} = 
A \delta^{\beta}_{\alpha} + B^{, \beta}_{, \alpha}, \eqno(175)
$$
where $F = C + \dot{D}, \; \psi = a^{2} (D/a^{2} \dot{)} + F, \; A = HF
- Y, \; B = - D/a^{2}$. So, the most general definition of $q$-scalar
in terms of 3-potentials is the following (cf. eq. (34)):
$$
q = A - 2Hv. \eqno(176)
$$

For the orthogonal gauge $(\psi = 0)$:
$$
C = 2a(aB \dot{)}, \;\;\; D = - a^{2} B, \;\;\; F = a^{2}\dot{B}.\eqno(177) 
$$ 
The next examples specify the function $B$ in eqs. (177).

For the synchronous gauge $(h_{oo} = \psi = 0)$:
$$
\dot{B} = a^{-2} Q - a^{-3} P, \;\;\; Q = \int \gamma q dt. \eqno(178a)
$$

For the comoving gauge $(v =\psi = 0)$:
$$
\dot{B} = \frac{q}{H a^{2}} - a^{-3} P. \eqno(178b)
$$

For the Newtonian gauge $B = \psi = 0$ (cf. eqs.(51, 52, 53)).

\section*{Appendix C}
One can easily verify that eqs. (42) are just the linear expansion
terms of the exact solution
$$
ds^{2} = dt^{2} - a^{2} \exp(2 \tau a_{\alpha \beta}) dx^{\alpha}
dx^{\beta}, \;\;\; \tau = \int \frac {dt}{a^{3}}, \eqno(179)
$$
where the function $a = a(t)$ can be found as follows:
$$
H^{2} = \frac{1}{3} \epsilon - \frac{1}{9a^6} \Lambda^2, \;\;\;
\Lambda^{2} = \frac{2}{3} a^{\beta}_{\alpha} a^{\alpha}_{\beta}, 
$$
$$ 
\dot H=-\frac 12(\epsilon+p)+\frac 13\frac{\Lambda^2}{a^6}.
$$

For $t \rightarrow 0$, we have the Kazner asymptotic:
$$
a^3=\Lambda t,\;\;\; g_{\alpha\beta}\sim{\rm diag}(t^{2p_a}),\eqno(180) 
$$
where Kazner exponents $(p_{1} + p_{2} + p_{3} = p^{2}_{1} + p^{2}_{2}
+ p^{2}_{3} = 1)$ are obviously related to the eigen values of matrix
$a_{\alpha \beta}$:
$$
det(a_{\alpha \beta} - \lambda \delta_{\alpha \beta}) = 0, \;\;\;\;
\lambda_{a} = \Lambda \left(p_{a} - \frac 13\right).
$$

Eqs. (170) describe solution for the Bianchi type I model with comoving
space. The infinite scale vortex and gravitational-wave perturbations
lead also to eqs. (179). Note, that the infinite scale perturbations 
although causing the expansion anisotropy (shear), do not perturb the 
spatial curvature and density perturbations $(\delta \epsilon = u_{\alpha}
= 0)$. It happens only in spatially flat Friedmann models. 

\newpage
\section*{References}
Birrel N. $\&$ Davies P.: 1981, {\it Quantum Fields In Curved Space}, 
Cambridge Univ. Press.\\
Colb E.W. $\&$ Turner M.S.: 1989, {\it The Early Universe}, Addison-Wesley 
Publ. C., INC.\\
Grib A.A., Mamaev S.G. $\&$ Mostepanenko A.: 1980, {\it Quantum Effects In 
Intensive Ex-\\ \hspace*{1cm} ternal Fields}, Atomizdat, Moscow.\\
Grichshuk L.P.: 1974, {\it Zh. Eksp. Teor. Fiz. (JETP)} {\bf 67}, 825.\\
Hodges H.M. $\&$ Blumenthal G.R.: 1989, {\it Inflation and Primordial 
Fluctuation Spectrum},\\ \hspace*{1cm} Preprint SCIPP 89/56.\\
Landau L.D. $\&$ Lifshitz E.M.: 1967, {\it Field Theory}, Nauka, Moscow.\\
Lifshitz E.M.: 1946, {\it Zh. Eksp. Teor. Fiz. (JETP)} {\bf 16}, 587.\\
Lifshitz E.M. $\&$ Pitaevski L.P.: 1978, {\it Statistical Physics}, Nauka, 
Moscow.\\
Linde A.D.: 1983, {\it Phys, Lett. B} {\bf 129}, 177.\\
Linde A.D.: 1986, {\it Phys. Lett. B} {\bf 175}, 395.\\
N.Lukash V.N.: 1980, {\it Zh. Eksp. Teor. Fis. (JETP)} {\bf 79}, 1601.\\
Lukash V.N. $\&$ Novikov I.D.: 1992, in {\it Observational and Physical 
Cosmology}, eds. F.Sanchez\\ \hspace*{1cm} et al., Cambridge Univ. Press, 3.\\
\end{document}